\documentclass{article}

\usepackage{microtype}
\usepackage{graphicx}
\usepackage{subcaption}

\usepackage{booktabs} 

\usepackage{hyperref}

\usepackage[accepted]{icml2024}

\usepackage{amsmath}
\usepackage{amssymb}

\usepackage{multirow}
\usepackage{multicol}
\usepackage{mathtools}
\usepackage{amsthm}

\usepackage{xspace}
\usepackage[capitalize,noabbrev]{cleveref}

\theoremstyle{plain}

\usepackage{subcaption}
\theoremstyle{definition}

\theoremstyle{remark}

\usepackage{inconsolata}
\usepackage{xspace}
\newcommand{\xxx}{CodeCoT\xspace}
\usepackage[capitalize]{cleveref}
\crefname{section}{Sec.}{Secs.}
\Crefname{section}{Section}{Sections}
\Crefname{table}{Table}{Tables}
\crefname{table}{Tab.}{Tabs.}
\definecolor{codegreen}{rgb}{0,0.6,0}
\definecolor{codegray}{rgb}{0.5,0.5,0.5}
\definecolor{codepurple}{rgb}{0.58,0,0.82}
\definecolor{backcolour}{rgb}{0.95,0.95,0.92}
\usepackage{listings}
\lstdefinestyle{mystyle}{
  backgroundcolor=\color{backcolour},   commentstyle=\color{codegreen},
  keywordstyle=\color{magenta},
  numberstyle=\tiny\color{codegray},
  stringstyle=\color{codepurple},
  basicstyle=\tiny,
  breakatwhitespace=false,         
  breaklines=true,                 
  captionpos=b,                    
  keepspaces=true,                 
  numbers=left,                    
  numbersep=10pt,                  
  showspaces=false,                
  showstringspaces=false,
  showtabs=false,                  
  tabsize=1
}

\usepackage[textsize=tiny]{todonotes}

\begin{document}

\twocolumn[
\icmltitle{CodeCoT: Tackling Code Syntax Errors in CoT Reasoning for Code Generation}



\icmlsetsymbol{equal}{*}

\begin{icmlauthorlist}
\icmlauthor{Dong Huang$^{*}$}{yyy}
\icmlauthor{Qingwen Bu$^{*}$}{xxx}
\icmlauthor{Yuhao Qing}{yyy}
\icmlauthor{Heming Cui}{yyy}
\end{icmlauthorlist}

\icmlaffiliation{yyy}{The University of Hong Kong, \{dhuang, yhqing, heming\}@cs.hku.hk}
\icmlaffiliation{xxx}{Shanghai AI Laboratory, qwbu01@sjtu.edu.cn}

\icmlcorrespondingauthor{Yuhao Qing}{yhqing@cs.hku.hk}

\icmlkeywords{Machine Learning, ICML}

\vskip 0.3in
]

\printAffiliationsAndNotice{\icmlEqualContribution} 



\begin{abstract}
Chain-of-thought (CoT) has emerged as a groundbreaking tool in NLP, notably for its efficacy in complex reasoning tasks, such as mathematical proofs. However, its application in code generation faces a distinct challenge, i.e., although the code generated with CoT reasoning is logically correct, it faces the problem of syntax error~(e.g., invalid syntax error report) during code execution, which causes the CoT result's pass@1 in HumanEval even lower than the zero-shot result.

In this paper, we present Code Chain-of-Thought (CodeCoT) that integrates CoT with a self-examination process for code generation. CodeCoT begins with the LLMs using CoT for initial code development to ensure the generated code follows the correct logic flow. Then, CodeCoT will generate test cases to validate whether the code has syntax errors during the execution. CodeCoT then employs a self-examination phase, in which the generated code is executed against these test cases in the local environment. If the local environment raises error information~(e.g., invalid syntax error), CodeCoT will iteratively refine the code based on the feedback information. Within this loop, CodeCoT can make sure their generated codes not only follow the logic flow of the code description, but the syntax error will also be addressed with the self-examination process. Our evaluation results reveal that CodeCoT improves the effectiveness of code generation. For example, CodeCoT increases pass@1 from 75.6\% to 79.3\% for the HumanEval dataset.
\end{abstract}

\section{Introduction}
\begin{figure*}
    \centering
    \includegraphics[width=1\linewidth]{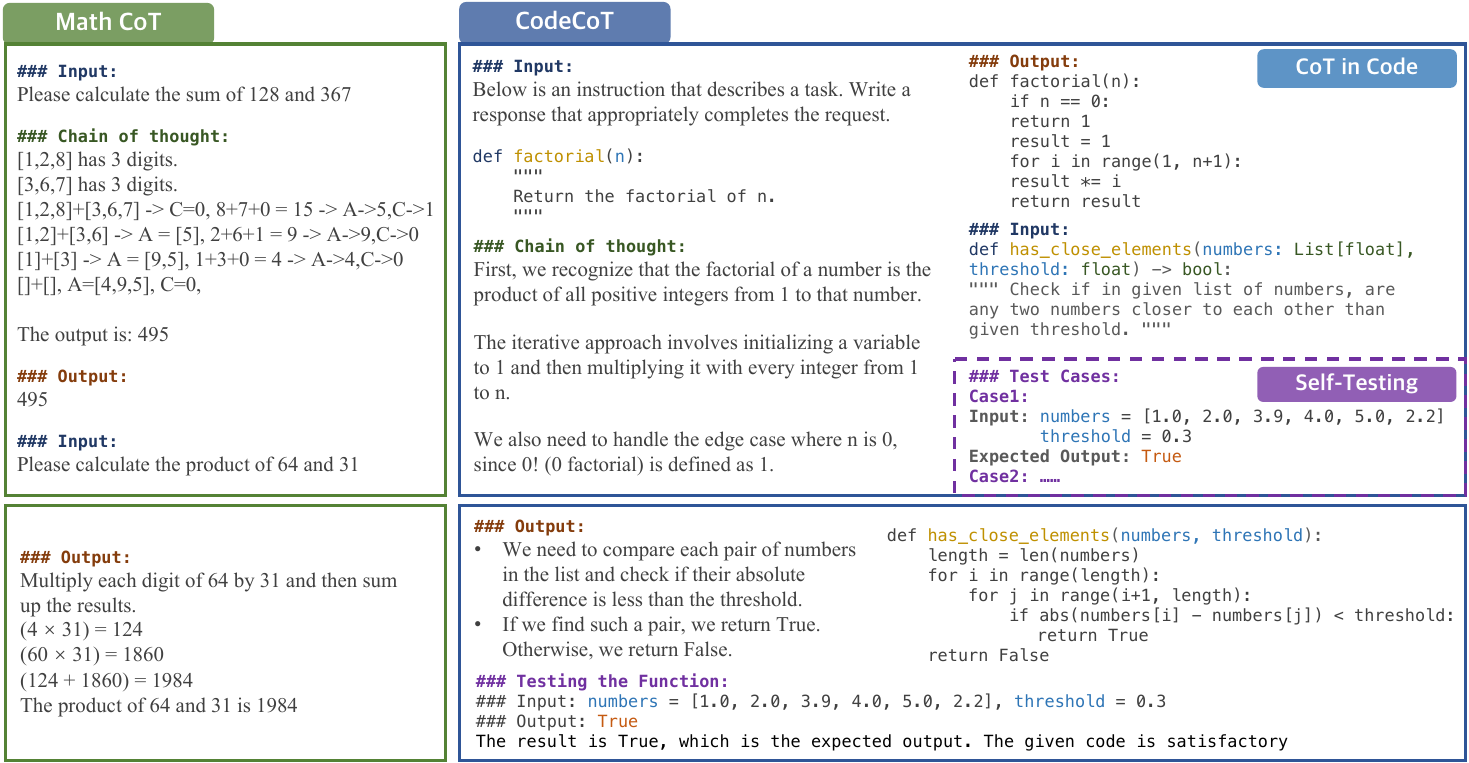}
    \caption{Illustration of Math Chain-of-Thought~(Math CoT), and CodeCoT.}
    \label{fig:intro}
\end{figure*}
Starting with the advances presented by the GPT-x models developed by OpenAI, transformer-based large language models (LLMs) currently provide state-of-the-art performance in many of the standard NLP tasks. One of the latest LLMs, GPT-3~\cite{brown2020language} uses about 175 billion parameters and was trained on an extremely large natural language training corpus, consisting of, among other things, excerpts from Wikipedia. Inspired by GPT-3, many large language models have been developed, which are different variants of the transformer architecture. Some of the most powerful models are PaLM~\cite{Chowdhery2022PaLMSL}, GLaM~\cite{Du2021GLaMES}, MegatronTuring NLG~\cite{Smith2022UsingDA}, Meta-OPT~\cite{Zhang2022OPTOP}, Gopher~\cite{Rae2021ScalingLM}, LaMDA~\cite{Thoppilan2022LaMDALM}, Chinchilla~\cite{Hoffmann2022TrainingCL}, ChatGPT, and GPT4~\cite{OpenAI2023GPT4TR}. GPT-4 currently provides state-of-the-art performance in NLP tasks such as natural language translation~\cite{Li2023ElicitingTT} and even translation to structured representations~\cite{Olmo2021GPT3toplanEP}.


Recently, \citet{Wei2022ChainOT} introduced the innovative concept of Chain-of-Thought Prompting. This technique guides a language model to produce a sequence of concise sentences that mirror the cognitive steps that a human might take when addressing a problem. As illustrated in Fig.\ref{fig:intro} \textbf{Math CoT}, when the user asks for the arithmetic query ``please calculate the sum of 128 and 367'', rather than directly answering with ``495'', the model using CoT would be driven to answer the question through the entire reasoning process. 
This method has demonstrated a marked improvement in model performance for various multistep reasoning challenges. The advent of CoT has unlocked new potential for LLM, especially in downstream tasks. CoT not only has improved performance in arithmetic, commonsense, and symbolic reasoning tasks~\cite{Zelikman2022STaRBR,Li2022MakingLM} but has also paved the way for more general innovations. These range from generating dialogues that capture user intentions to participating in multimodal tasks that fuse textual and visual information~\cite{liu-etal-2022-multi}. The growing interest in CoT underscores its potential to shape how LLMs interact and reason.

The wide use of CoT in math reasoning inspires us to discuss the application of CoT in code generation. However, as mentioned by \citet{Dong2023SelfcollaborationCG, Wang2023INTERVENORPT, Shinn2023ReflexionLA}, directly applying CoT in code generation will decrease code generation effectiveness, exemplified by lower pass@1 scores than zero-shot results. As shown in \cref{fig:intro} \textbf{CoT in Code}, we observe that the key reason is code generated with CoT reasoning although follows the logic flow~(e.g., pseudocode) of task description, it ignores to follow the syntax requirements (e.g., avoiding syntax errors), which then decreases the pass@1 of CoT reasoning in code generation tasks~(e.g., HumanEval and MBPP dataset).

To address the challenge of CoT in code generation, we propose CodeCoT, a novel framework that incorporates CoT reasoning and self-examination to mitigate the issues arising from the disconnect between narrative reasoning and strict code syntax requirements. 
During the code generation process, CodeCoT first utilizes CoT reasoning to generate code with the correct logic flow and then utilizes the self-examination process to detect and fix the syntax errors in its code. This dual-focused approach effectively bridges the gap observed in traditional CoT applications for code generation. As shown in Fig~\ref{fig:intro} \textbf{CodeCoT}, we distinguish the CodeCoT framework into the \textit{CoT in Code} and the \textit{self-examination}. The \textit{CoT in Code} aligns closely with other CoT techniques used in various downstream tasks of LLMs. It contains a task description, CoT reasoning, task output, and another task description. The \textit{self-examination} component will introduce a new layer of self-examination into the process. Specifically, for the \textit{self-examination} component, the LLM will generate a few test cases. Then the \textit{self-examination} component will execute the code with these test cases in the local environment to assess whether syntax errors exist in the code. If the execution raises a syntax error, the LLM will then regenerate the code function to iterate and refine the code accordingly. This process ends with the LLM producing a polished code that is not only logic correct but also does not contain syntax errors.

Extensive experiments illustrate that CodeCoT showed a notable increase in pass@1 accuracy for the evaluation datasets. For example, CodeCoT increases pass@1 from 75.6\% and 69.8\% to 79.3\% and 89.5\% for HumanEval and MBPP datasets. CodeCoT also obtains SOTA performance in HumanEval-ET and MBPP-ET datasets. For example, CodeCoT obtains 69.5\% and 63.0\% pass@1 while baselines only obtain 56.1\% and 49.5\% pass@1 in HumanEval-ET and MBPP-ET. 
Our main contributions are as follows:

\begin{itemize}
\item We propose CodeCoT, which utilizes a self-examination process to address the challenge of Chain-of-Thought~(CoT) in code generation~(i.e., transforming narrative-based logical reasoning into precise and executable code).
\item CodeCoT obtains SOTA performance on the HuamEval benchmark and significantly improves pass@1 accuracy compared to existing methods. For example, CodeCoT increases pass@1 from 75.6\% and 69.8\% to 79.3\% and 89.5\% for HumanEval and MBPP datasets.
\end{itemize}

\section{Related Work}

\subsection{Large language models}
The trajectory of language model development has witnessed a consistent emphasis on scaling, both in terms of the model architecture and the datasets they are trained on. This chronology of growth can be traced back to the works of~\citet{brants-etal-2007-large}, who demonstrated the advantages of models trained on a colossal 2 trillion tokens, resulting in the generation of 300 billion n-grams. This substantial leap was especially pertinent to enhancing machine translation quality. Even though the early techniques, such as the ``Stupid Backoff'' for smoothing, were rudimentary, advancements were made by~\citet{heafield-etal-2013-scalable}. The transformative potential of scaling was further emphasized with the evolution of transformer architectures, which carved out novel benchmarks in numerous NLP challenges. Some trailblazing models in this era include BERT by~\citet{devlin-etal-2019-bert}, GPT-2 by~\citet{Radford2019LanguageMA}, MegatronLM by~\citet{Shoeybi2019MegatronLMTM}, and T5 by~\citet{Raffel2019ExploringTL}. The landscape experienced a monumental shift with the introduction of GPT-3 by~\citet{Brown2020LanguageMA}, a behemoth with 175 billion parameters. This spurred the development of a lineage of Large Language Models such as Jurassic-1, Megatron-Turing NLG, Gopher, Chinchilla, PaLM, OPT, and GLM, introduced between 2021 and 2022. Delving into the mechanics of scaling, studies like that by~\citet {Hestness2017DeepLS} and~\citet{Rosenfeld2019ACP} evaluated the relationship between model and dataset sizes and resultant performance, unearthing the presence of power laws.

\subsection{Chain of Thought Prompting}
The concept of chain-of-thought prompting was introduced to harness the reasoning capabilities of large language models, presenting a novel approach to refining the performance of these models in intricate tasks. Initially proposed by \cite{Wei2022ChainOT}, this technique aimed to supplement few-shot examples with detailed reasoning steps, leading to significant performance enhancements in complex tasks. Over time, this approach inspired a plethora of research, each seeking to fine-tune and enhance the foundational concept of CoT. Noteworthy advancements include innovations in self-consistency \cite{Wang2022SelfConsistencyIC}, advancements in least-to-most prompting and its dynamic variant \cite{Zhou2022LeasttoMostPE, Zhou2023EfficientPV}, as well as breakthroughs in bootstrapping \cite{Zelikman2022STaRBR} and self-training \cite{Huang2022LargeLM}. The verifier methodology \cite{Li2022MakingLM} also stands out as a remarkable contribution in this realm. A notable exception in the landscape of adaptability is Auto-CoT \cite{Zhang2022AutomaticCO}. This method categorizes test questions into distinct clusters to enhance diversity, subsequently generating answers through zero-shot prompting.

\begin{figure*}
    \centering
    \includegraphics[width=\linewidth]{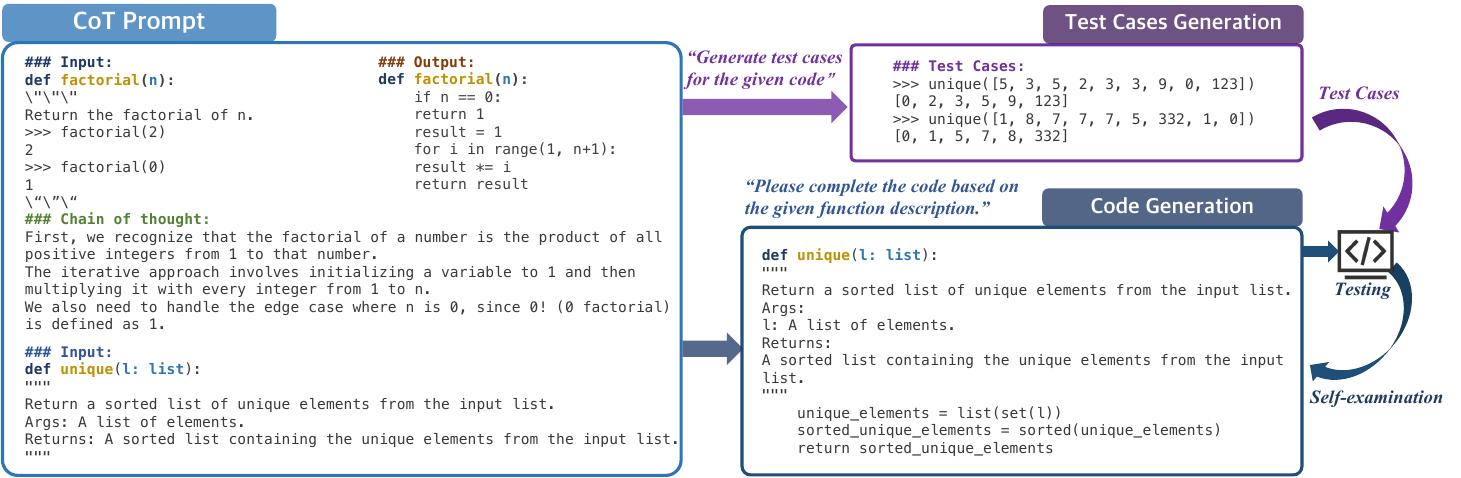}
    \caption{CodeCoT's workflow. It contains four components, i.e., CoT Prompt, Test Cases Generation, which is used to generate test cases for the given tasks, Code Generation, and Self-examination~(testing with self-correction).}
    \label{fig:pipeline}
\end{figure*}

\subsection{Chain of thought application in LLMs}
Following the initial chain of thought prompting proposed by~\citet{Wei2022ChainOT} used in arithmetic, commonsense and symbolic reasoning, lots of works spring up aim to improve different parts of original reasoning processing, including autocot~\cite{Zhang2022AutomaticCO}, self-consistency~\cite{Wang2022SelfConsistencyIC}, active prompt~\cite{Diao2023ActivePW}, automate-cot~\cite{Shum2023AutomaticPA}. Besides that, there are some pioneers who apply similar ideas to knowledgeable dialogue generation and other tasks. ~\cite{liu-etal-2022-multi} utilizes a multi-stage prompting approach to generate knowledge first and then response, achieving better performance than fine-tuning in terms of response knowledgeability and engagement. ~\cite{Tan2021MSPMP} fuse the prompting with the tuning to shift the pre-trained models to translation tasks. The recent related work converts user-profiles and historical iterations into prompts to build conversational recommender systems with the backbone as ChatGPT~\cite{Gao2023ChatRECTI}.~\citet{Cobbe2021TrainingVT} employ a calculator for arithmetic operations as a post hoc processing, and~\citet{demeter-etal-2020-stolen} add specialized modules for generating cities and dates. Unlike these works, PAL generates code for a Python interpreter, which is general enough to handle both arithmetic calculations and dates, without specialized modules and ad-hoc fixes.~\citet{Chowdhery2022PaLMSL} has also experimented with external calculators.~\citet{Pi2022ReasoningLP} pretrain the model on execution results of random expressions on a calculator.

\subsection{Enhancing Code Generation through Prompt Engineering}
\citep{Chen2021EvaluatingLL} introduced a simple filtering approach by selecting only output samples that successfully pass the public test cases. AlphaCode \cite{Li2022CompetitionlevelCG}, CodeT \cite{Chen2022CodeTCG}, and MBR-Exec \cite{Shi2022NaturalLT} proposed to generate more test cases and use more sophisticated rule-based methods to rank generation samples by their execution behaviors. LEVER \cite{Ni2023LEVERLT}, Coder-Reviewer \cite{Zhang2022CoderRR} and Code Rankers \cite{Inala2022FaultAwareNC} follow a similar principle but introduce more model-based ranking methods. Recently, more related works have been proposed to boost generation quality through iterative self-revisions. Self-Edit \cite{Zhang2023SelfEditFC} utilizes test outcomes from public test cases as feedback for models to self-revise their codes. Self-correct \cite{Welleck2022GeneratingSB} and CodeRL \cite{Le2022CodeRLMC} introduce secondary models to predict the correctness of output programs and revise them accordingly. Self-debug \cite{Chen2023TeachingLL}, Sef-refine \cite{Madaan2023SelfRefineIR}, and Reflexion \cite{Shinn2023ReflexionLA} propose to facilitate better code revision with synthetic natural language explanation or reflection self-generated by LLMs. Self-repair \cite{Olausson2023IsSA} and ILF \cite{Chen2023ImprovingCG} follow a similar strategy but highlight the use of natural language explanation provided by human experts. 

Recently, some CoT strategies which were proposed in parallel with CodeCoT, have been employed to enhance the effectiveness of code generation. For example, \citep{Li2023StructuredCP} proposes SCOT  to utilize CoT reasoning for structure-level code generation. CodeChain \cite{Le2023CodeChainTM} extends SCOT and generates code from a more fine-grained sub-module level compared with SCOT. Different from these parallel CoT strategies, which focus on the sub-module/structure level code generation with CoT reasoning, CodeCoT rethinks the challenge of CoT reasoning-guided code generation, i.e., current CoT reasoning addresses the logic requirement of code generation while ignoring the syntax requirements at the same time. CodeCoT addresses the above challenge by utilizing CoT reasoning and the self-examination process to obtain logic and syntax correctness during the code generation process.

\section{Methodology}

\subsection{Overview}
\cref{fig:pipeline} provides the pipeline of CodeCoT, divided into four pivotal components: the CoT Prompt, Test Cases Generation, Code Generation, and Self-Examination. The process initiates with the CoT Prompt, where the LLM is given a task, for example, to ``calculate the factorial of n.'' The LLM then breaks down this task, providing a logical chain of thought detailing its approach to tackle the task, resulting in a generated code function. Subsequently, in the Test Cases Generation phase, the LLM will generate a set of tests used to evaluate whether the code is executable. In the Code Function Generation phase, the LLM will first generate with correct logic flow. Then it will based on the feedback information refine the code to fix the syntax errors. During the Self-Examination stage, the code will be executed in the local environment with test cases to analyze whether the code has bugs~(e.g., invalid syntax error). If errors arise during the local execution process, the error information~(feedback information) will be fed into the LLM to refine the code, to ensure the code has both logic flow and syntax correct. The iterative procedure of self-examination will be conducted through a series of multi-step iterations, allowing user-specifiable iteration quantities, while the default is five iterations.

\subsection{CoT in Code}
\paragraph{CodeCoT prompt.} 
The first stage in the CodeCoT process is providing the model with a clear and structured prompt, as shown in Fig~\ref{fig:pipeline}. CoT prompt contains two key elements, i.e., task description and example for guidance. The task description will have a concise statement that outlines the main objective of the code to be generated. Using the given example, the task could be ``Return a sorted list of unique elements from the input list.'' This sets a clear goal for the code generation process and guides the LLM in its task. The example for guidance provides a code generation example to serve as a guiding post for the LLM. In this instance, the example given is ``implement the factorial function.'' While the example may not directly correlate with the main task, it provides insight into the kind of logic or structure expected in the solution. This example aids the model in understanding the depth, complexity, or approach the user anticipates for the task at hand.

\begin{figure}[h]
    \centering
    \includegraphics[width=\linewidth]{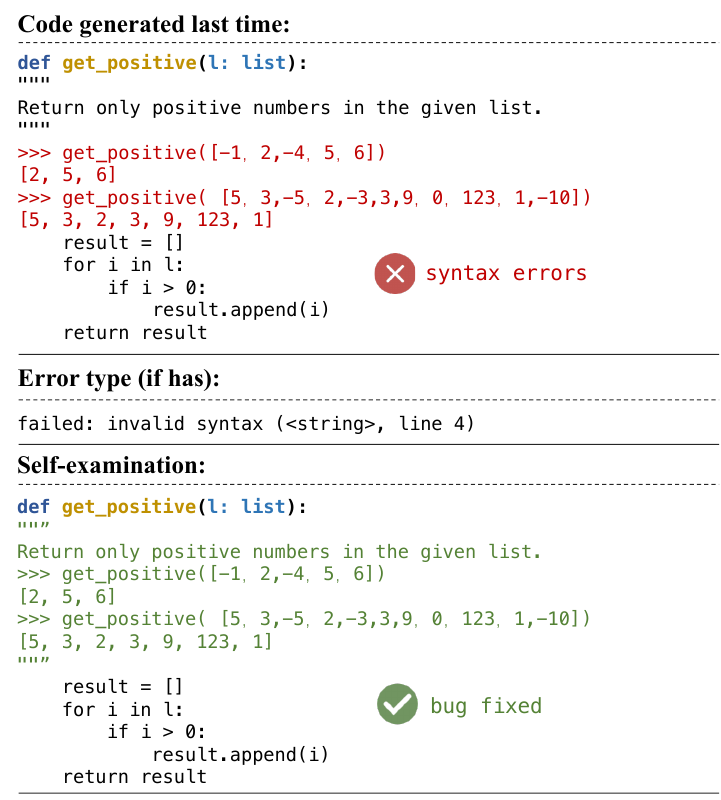}
    \caption{An illustration of Self-exam CodeCoT addresses bugs in their generated code functions. The LLM will first call the terminal and then evaluate the generated code function with its generated tests, if the terminal raises an error, the LLM will then revise the code based on the error information.}
    \label{fig:selfexam}
\end{figure}

\paragraph{Code Function Generation.}
After receiving a prompt that includes both a task example and a task description, the LLM will then generate code, primarily guided by the task description. This initial code logic flow will be generated by Chain of Thought (CoT) reasoning, ensuring that the logical flow of the code aligns with the problem-solving process. Once the generated code contain syntax errors, a self-examination mechanism is activated, identifying these errors within a local environment. This error feedback is then feedback to the LLM, enabling it to refine and correct the syntax. This iterative process of self-examination and feedback ensures the code not only logically correct but also syntactically accurate.

\subsection{CodeCoT}
\paragraph{Test case generation.} To evaluate code generated by the CoT prompt in the local environment, we will also require the LLM to generate test cases that are used to evaluate the syntax of the generated code. Specifically, as shown in~\cref{fig:pipeline}, during the code generation procedure, we will also require the LLM \textit{generate test cases for the given code}\footnote{We commonly require the LLM to generate five test cases for the code since the test cases provide by dataset on average lower than 5.}. To reduce the overhead of API communication, we require the code generation and test cases generation in the same prompt~(see~\cref{fig:codegenerationprompt}).

\paragraph{Self-examination.}
The CoT prompt makes the LLMs generated code sometimes logically correct, but the code sometimes contains syntax errors, which cause the code to not be executed correctly. As shown in~\cref{fig:pipeline} self-examination with test cases, we address the above problem by providing a self-examination stage that employs a continuous examination and iterative refinement to address this problem. Specifically, once the LLM produces an initial draft of the code function, it is subject to rigorous scrutiny with its self-generated tests by executing the code with tests in the local environment ~(local terminal). Then CodeCoT will obtain the feedback from the local environment. If the error messages, e.g. invalid syntax error, arise during this phase, CodeCoT will revise the code based on the last generated code and feedback from the local environment.  For example, as illustrated in~\cref{fig:selfexam}, we can observe that the above code function~(first version) has a syntax error in line 4. The key reason is that the triple quotation mark should be in the 8th line. We can find that although the code follows the CoT reasoning, the syntax error causes the code not to execute, so Self-exam CodeCoT will feed the reported error to the LLM, it then revises this error and reports a correct version in the below. The revised function is then re-tested. Once the revised function passes all tests, we can then consider it functionally reliable and syntactically sound. 

\section{Evaluation}

In this section, we evaluate \xxx to answer the following questions:
\begin{itemize}
    \item RQ1. How does CodeCoT perform?
    \item RQ2. Does CodeCoT fix syntax errors during the self-examination process?
    \item RQ3. How do self-examination steps affect CodeCoT's effectiveness?
    \item RQ4. How does CodeCoT each component perform?
\end{itemize}

We use pass@1 as the evaluation metric for code correctness, the most widely adopted metric in the literature of automatic code generation~\cite{Chen2021EvaluatingLL,Austin2021ProgramSW,Dong2023CodeScoreEC,Zhang2023SelfEditFC,Dong2023SelfcollaborationCG}.
\begin{table*}[h]
    \centering
    \caption{End-to-end results of CodeCoT and baseline approaches for HumanEval, MBPP, and their ET datasets. ``-'' means the technique does not report the results and we can not reproduce results due to lack of source code and API.}
    \begin{tabular}{l|cccc}
    \toprule
    Models&HumanEval&HumanEval-ET&MBPP&MBPP-ET\\
    \midrule
    \textbf{Customize model}\\
    \midrule
         AlphaCode~(1.1B)&17.1&-&-&-  \\
         Incoder~(6.7B)&15.2&11.6&17.6&14.3 \\
         CodeGeeX~(13B)&18.9&15.2&26.9&20.4\\
         StarCoder~(15.5B)&34.1&25.6&43.6&33.4\\
         CodeGen-Mono~(16.1B)&32.9&25.0&38.6&31.6\\
         CodeX~(175B)&47.0&31.7&58.1&38.8\\
         CodeX~(175B)+CodeT&65.8&51.7&67.7&45.1\\
         ChatGPT&57.3&42.7&52.2&36.8\\
         GPT-4&67.6&50.6&68.3&52.2\\
         \midrule
         \textbf{ChatGPT with Prompting}\\
         \midrule
         Few-Shot&67.7&54.9&65.8&48.3\\
         ReAct&56.9&49.4&67.0&45.9\\
         Reflexion&68.1&50.6&70.0&47.5\\
         ToT&54.4&42.7&65.8&40.8\\
         RAP&63.1&52.4&71.4&46.7\\
         Self-Edit&62.2&54.3&56.4&45.9\\
         Self-Planing&65.2&48.8&58.6&41.5\\
         Self-Debugging&61.6&45.8&60.1&52.3\\
         Self-Collaboration&74.4&56.1&68.2&49.5\\
         INTERVENOR&75.6&54.8&69.8&47.1\\
         SCOT&60.6&-&47.0&-\\
         CodeChain&62.8&54.3&59.1&45.5\\
         Vanilla CodeCoT&69.5&58.5&67.7&48.6\\
        CodeCoT&\textbf{79.3}&\textbf{69.5}&\textbf{89.5}&\textbf{63.0}\\
         \bottomrule
    \end{tabular}
    \label{tab:end2end}
\end{table*}

\begin{table*}
    \centering
    \caption{Evaluation results of error type distribution in pass@1 evaluation for HumanEval and MBPP datasets. We classify the Non-AssertError that exists since the code does not pass the tests as SyntaxErrors.}
    \begin{tabular}{c|ccccccc}
         \toprule
         Strategies&\multicolumn{2}{c}{HumanEval}&\multicolumn{2}{c}{MBPP}\\
         &AssertError&SyntaxError&AssertError&SyntaxError\\
         \midrule
         CoT&64\%&36\%&67\%&33\%\\
         SCOT&70\%&30\%&68\%&32\%\\
         CodeChain&65\%&35\%&66\%&34\%\\
         1 step~(CodeCoT)&73\%&27\%&75\%&25\% \\
         3 step~(CodeCoT)&86\%&14\%&87\%&13\% \\
         5 step~(CodeCoT)&98\%&2\%&99\%&1\% \\
         \bottomrule
    \end{tabular}
    \label{tab:assert_distribution}
\end{table*}

\paragraph{Datasets.} In this paper, we evaluate \xxx's effectiveness with four widely used code generation datasets, i.e., HumanEval~\cite{Chen2021EvaluatingLL} and MBPP~\cite{Austin2021ProgramSW}, and their enhanced versions, i.e., HumanEval-ET and MBPP-ET~\cite{Dong2023CodeScoreEC}. HumanEval and HumanEval-ET focus on a range of programming challenges, offering a diverse set of problems to test the model's problem-solving skills and adaptability. On the other hand, MBPP and MBPP-ET provide a comprehensive collection of Python programming problems, designed to evaluate the model's proficiency in Python syntax and its ability to handle a variety of coding scenarios. The enhanced versions, HumanEval-ET and MBPP-ET, include more adequate test cases, making them more challenging and better suited for evaluating advanced models.

\paragraph{Baselines} 
To illustrate the effectiveness of \xxx, in this paper, we compare \xxx with several large language models (LLMs), including both open-source and closed-source models, such as AlphaCode~\cite{Li2022CompetitionlevelCG}, Incoder~\cite{Fried2022InCoderAG}, CodeGeeX~\cite{Zheng2023CodeGeeXAP}, StarCoder~\cite{Li2023StarCoderMT}, CodeGen-Mono~\cite{Nijkamp2022CodeGenAO}, CodeX, CodeX with CodeT~\cite{Chen2022CodeTCG}, ChatGPT, and GPT4~\cite{OpenAI2023GPT4TR}. Furthermore, we evaluated \xxx with current SOTA prompt engineering methods, i.e., Few-shot, ReAct~\cite{Yao2022ReActSR}, Reflexion~\cite{Shinn2023ReflexionLA}, ToT~\cite{Yao2023TreeOT}, RAP~\cite{Hao2023ReasoningWL}, Self-Edit~\cite{Zhang2023SelfEditFC}, Self-Planing~\cite{Jiang2023SelfplanningCG}, Self-Debugging~\cite{Chen2023TeachingLL}, Self-Collaboration~\cite{Dong2023SelfcollaborationCG}, SCOT~\cite{Li2023StructuredCP},
CodeChain~\cite{Le2023CodeChainTM}, 
and INTERVENOR~\cite{Wang2023INTERVENORPT}. The base model used in our prompt engineering strategies is ChatGPT.
These strategies have been shown to significantly improve the performance of LLMs in complex code generation scenarios~\footnote{Although we compare with CodeChain and INTERVENOR, we should clarify CodeChain and INTERVENOR is later proposed compared with of \xxx.}.

\begin{table*}
    \centering
\caption{Evaluation results of CodeCoT with different refine steps.}
    \begin{tabular}{c|cccc}
    \toprule
    Step&HumanEval&HumanEval-ET&MBPP&MBPP-ET\\
    \midrule
         1&71.3&60.4&81.7&58.4\\
         2&73.8&64.6&86.8&61.5\\
         3&76.8&67.1&88.3&62.7\\
         4&78.7&68.9&89.1&62.7\\
         5&79.3&69.5&89.5&63.0\\
         \bottomrule
    \end{tabular}
    \label{tab:steps}
\end{table*}

\subsection{RQ1. How does CodeCoT perform?}

The evaluation results of CodeCoT and the baselines are shown in~\cref{tab:end2end}, where we can find that \xxx achieves SOTA performance compared to baseline models and prompt engineering strategies in the HumanEval and MBPP datasets. For example, we can find that ChatGPT obtains 57.3\% and 52.2\% pass@1 in HumanEval and MBPP datasets. While CodeCoT obtains 79.3\% and 89.5\% pass@1 in these datasets, increasing 22\% and 37.3\% pass@1 in HumanEval and MBPP datasets, which illustrates that CodeCoT can improve the code generation effectiveness of its based model. Then, when we compare CodeCoT with baseline prompt engineering strategies, we can find that CodeCoT still obtains the SOTA performance. For example, compared with Self-Collaboration and INTERVENOR, CodeCoT improves the pass@1 from 74.4\% and 75.6\% to 79.3\% in the HumanEval dataset, and CodeCoT also improves the pass@1 from 68.2\% and 69.8\% to 89.5\% in MBPP dataset. For HumanEval-ET and MBPP-ET datasets, CodeCoT also increases the pass@1 from 56.1\% and 49.5\% to 69.5\% and 63.0\% compared with Self-Collaboration. Compared with current CoT results, we can also find that SCOT~\cite{Li2023StructuredCP} and CodeChain~\cite{Le2023CodeChainTM} are also lower than CodeCoT, which is due to the SCOT and CodeChain do not address the challenge of syntax errors in the code generation procedure.

\subsection{RQ2. Does CodeCoT fix syntax errors during the self-examination process?}
To illustrate whether CodeCoT decreases syntax error during the self-examination process compared with other CoT strategies, we further analyze the error distribution when we calculate pass@1 in the experiment. We divided the error types into AssertError~(which commonly existed in the pass@1 calculation process since code snippets do not pass the test cases.) and Other errors, which means the code can not be executed since syntax, compiler, and other errors that do not raise due to the assertion in the code. The evaluation results are shown in~\cref{tab:assert_distribution}, where we can observe that the ratio of RuntimeError~(e.g., SyntaxError) is lower than our baselines. For example, in SCOT and CodeChain, the RuntimeError has 30\% and 35\% in the HumanEval dataset, while 1-step CodeCoT only has 27\%, and when we increase the self-examination steps, the RuntimeError further decreases to 2\% for 5-step. These results illustrate that compared with other CoT reasoning strategies, CodeCoT can reduce the errors caused by syntax errors.

\subsection{How do self-examination steps affect CodeCoT's effectiveness?}

To evaluate the influence of iterative self-examination steps on code generation performance, we systematically increased the number of self-examination rounds and monitored the resulting accuracy improvements. The evaluation results are shown in~\cref{tab:steps}, where we can find that increasing the self-examination steps can improve code generation effectiveness. For example, when we increase the step from 1 to 5, the pass@1 of \xxx increases from 71.3\% and 81.7\% to 79.3\% and 89.5\% for HumanEval and MBPP datasets. These behaviors are also shown in the ET datasets. For example, pass@1 of \xxx also increases from 60.4\% and 58.4\% to 69.5\% and 63.0\% for HumanEval-ET and MBPP-ET datasets.

\begin{table*}
    \centering
    \caption{Evaluation results of how \xxx's component affects its effectiveness.}
    \begin{tabular}{c|cccc}
    \toprule
    Prompt&HumanEval&HumanEval-ET&MBPP&MBPP-ET\\
    \midrule
    Coder&67.7&54.9&65.8&48.3\\
    +CoT&69.5&58.5&67.7&48.6\\
    +Self-examination&70.1&57.9&79.0&56.4\\
    CodeCoT&79.3&69.5&89.5&63.0\\
        \bottomrule
    \end{tabular}
    \label{tab:component}
\end{table*}

\subsection{RQ4. How does CodeCoT each component perform?}
As shown in~\cref{fig:pipeline}, during the code generation procedure, \xxx will first utilize the CoT prompt to generate code and test cases, then execute in the local environment to analyze whether the code is correct and then refine the code if there was a run-time error. In this section, we will analyze how different components of \xxx affect its effectiveness. Specifically, we will compare the benign effectiveness of few-shot ChatGPT~(Coder), Coder + CoT prompt, Coder + Self-examination process~(generate test cases and execute in the local environment), Coder + CoT prompt + Self-examination~(CodeCoT).

The evaluation results are shown in~\cref{tab:component}. We can find that first, with the assistance of each component, e.g., the CoT prompt and the self-examination, the pass@1 will increase compared with the result of only the Coder. For example, once we use the Coder and CoT prompt, the pass@1 increases from 67.7\% and 65.8\% to 69.5\% and 67.7\% for HumanEval and MBPP datasets. Then, when we combine the  Coder+Self-examination, the pass@1 is further increased. For instance, the pass@1 increases from 67.7\% and 65.8\% to 70.1\% and 79.0\% for HumanEval and MBPP datasets. We can also observe that the pass@1 of the ET version is also increased, e.g., the pass@1 increases from 54.9\% and 48.3\% to 57.9\% and 56.4\% for HumanEval-ET and MBPP-ET datasets. Although Coder+CoT prompt and Coder+Self-examination improve the code generation effectiveness of ChatGPT, we can find that they do not obtain the SOTA performance compared with our baselines. For example, Slef-Collaboration obtains 74.4\% and 68.2\% pass@1 in HumanEval and MBPP datasets. However, once we combine three components into CodeCoT, we can find that the pass@1 further achieves 79.3\% and 89.5\% in HumanEval and MBPP datasets, which is higher than our baseline strategies, e.g., 75.6\% and 69.8\% pass@1 in HumanEval and MBPP, proving that in CodeCoT each component is important and cannot be ignored.

\begin{table}[h]
    \centering
    \caption{Evaluation for the test case effectiveness.}
    \begin{tabular}{c|cc}
    \toprule
    Models&HumanEval&MBPP\\
    \midrule
        Self-examination&47.0&57.2\\
        CodeCoT&67.1&79.0\\
         \bottomrule
    \end{tabular}
    \label{tab:testcase}
\end{table}

\subsection{Further discuss for the \xxx's effectiveness with its components}
In this section, we want to discuss why CodeCoT pass@1 will increase from 68.3\%/70.1\% and 73.9\%/79.0\% to 79.3\% and 89.5\% in~\cref{tab:component}. Specifically, we analyze the test case generated by Coder+Self-examination and CodeCoT in~\cref{tab:testcase}. We can find that for the Coder+Self-examination, the pass@1 of the test cases on \textit{canonical\_solution} only have 47.0\% and 57.2\% in HumanEval and MBPP. However, when the CoT prompt is included in CodeCoT, the pass@1 of test cases on \textit{canonical\_solution} increases to 67.1\% and 79.0\% in these datasets, which illustrates why CodeCoT obtains higher performance than Coder+Self-examination. Specifically, CodeCoT obtains more accuracies test cases for self-testing and these test cases can accurately guide the Coder to refine its generated code during self-examination procedure.

\section{Conclusion}
In this paper, we address the commonly existed runtime errors of utilizing the CoT prompt in code generation by proposing CodeCoT, which utilizes the self-examination procedure to detect runtime errors in the generated code with its generated tests. Our evaluations reveal that the proposed CodeCoT significantly improves the pass@1 of code generation across various LLMs.  For example, CodeCoT obtains 79.3\% and 89.5\% pass@1 in HumanEval and MBPP datasets for ChatGPT with the self-examination procedure. Future work could investigate further refinements to our approach, explore its applicability to other domains, and delve deeper into the underlying mechanisms that make CoT so effective for LLM-based code generation.

\section{Limitations}
We discuss the limitations of our work that could hopefully inspire future research in this avenue. First, in this paper, we focus on the close-sourced models, e.g., ChatGPT~(in most experiments), GPT-4~(in appendix). Open-sourced models, e.g., CodeT5+, StarCoder, CodeGen, and others are not evaluated in our paper. The key reason is that CoT requires large parameters~($>$175B) can illustrate its effectiveness~\cite{Wei2022ChainOT}, which constrain CodeCoT's application in the open-source model. Therefore, we encourage future investigation for other reasoning strategies~(e.g., ToT, GoT) in the open-source model with our self-examination steps. Second, \xxx requires multiple interactions with LLM for the self-examination procedure, which will increase the overhead of code generation both for time and for the API fee. So we encourage future studies to decrease the interaction times. Finally, as shown in~\cref{tab:testcase}, the tester in the \xxx can not make sure all test cases are correct, which means that some code snippets generated by the CodeCoT may then revises to an error version. In the future, we will try to quantify the revised functions and investigate how to avoid these behaviors.

\newpage
\bibliography{example_paper}

\begin{thebibliography}{65}
\providecommand{\natexlab}[1]{#1}
\providecommand{\url}[1]{\texttt{#1}}
\expandafter\ifx\csname urlstyle\endcsname\relax
  \providecommand{\doi}[1]{doi: #1}\else
  \providecommand{\doi}{doi: \begingroup \urlstyle{rm}\Url}\fi

\bibitem[Austin et~al.(2021)Austin, Odena, Nye, Bosma, Michalewski, Dohan, Jiang, Cai, Terry, Le, and Sutton]{Austin2021ProgramSW}
Austin, J., Odena, A., Nye, M., Bosma, M., Michalewski, H., Dohan, D., Jiang, E., Cai, C.~J., Terry, M., Le, Q.~V., and Sutton, C.
\newblock Program synthesis with large language models.
\newblock \emph{ArXiv}, abs/2108.07732, 2021.
\newblock URL \url{https://api.semanticscholar.org/CorpusID:237142385}.

\bibitem[Brants et~al.(2007)Brants, Popat, Xu, Och, and Dean]{brants-etal-2007-large}
Brants, T., Popat, A.~C., Xu, P., Och, F.~J., and Dean, J.
\newblock Large language models in machine translation.
\newblock In \emph{Proceedings of the 2007 Joint Conference on Empirical Methods in Natural Language Processing and Computational Natural Language Learning ({EMNLP}-{C}o{NLL})}, pp.\  858--867, Prague, Czech Republic, June 2007. Association for Computational Linguistics.
\newblock URL \url{https://aclanthology.org/D07-1090}.

\bibitem[Brown et~al.(2020{\natexlab{a}})Brown, Mann, Ryder, Subbiah, Kaplan, Dhariwal, Neelakantan, Shyam, Sastry, Askell, Agarwal, Herbert-Voss, Krueger, Henighan, Child, Ramesh, Ziegler, Wu, Winter, Hesse, Chen, Sigler, Litwin, Gray, Chess, Clark, Berner, McCandlish, Radford, Sutskever, and Amodei]{Brown2020LanguageMA}
Brown, T.~B., Mann, B., Ryder, N., Subbiah, M., Kaplan, J., Dhariwal, P., Neelakantan, A., Shyam, P., Sastry, G., Askell, A., Agarwal, S., Herbert-Voss, A., Krueger, G., Henighan, T.~J., Child, R., Ramesh, A., Ziegler, D.~M., Wu, J., Winter, C., Hesse, C., Chen, M., Sigler, E., Litwin, M., Gray, S., Chess, B., Clark, J., Berner, C., McCandlish, S., Radford, A., Sutskever, I., and Amodei, D.
\newblock Language models are few-shot learners.
\newblock \emph{ArXiv}, abs/2005.14165, 2020{\natexlab{a}}.
\newblock URL \url{https://api.semanticscholar.org/CorpusID:218971783}.

\bibitem[Brown et~al.(2020{\natexlab{b}})Brown, Mann, Ryder, Subbiah, Kaplan, Dhariwal, Neelakantan, Shyam, Sastry, Askell, et~al.]{brown2020language}
Brown, T.~B., Mann, B., Ryder, N., Subbiah, M., Kaplan, J., Dhariwal, P., Neelakantan, A., Shyam, P., Sastry, G., Askell, A., et~al.
\newblock Language models are few-shot learners.
\newblock \emph{arXiv preprint arXiv:2005.14165}, 2020{\natexlab{b}}.

\bibitem[Chen(2023)]{Chen2023ImprovingCG}
Chen, A.
\newblock Improving code generation by training with natural language feedback.
\newblock \emph{ArXiv}, abs/2303.16749, 2023.
\newblock URL \url{https://api.semanticscholar.org/CorpusID:257804798}.

\bibitem[Chen et~al.(2022)Chen, Zhang, Nguyen, Zan, Lin, Lou, and Chen]{Chen2022CodeTCG}
Chen, B., Zhang, F., Nguyen, A., Zan, D., Lin, Z., Lou, J.-G., and Chen, W.
\newblock Codet: Code generation with generated tests.
\newblock \emph{ArXiv}, abs/2207.10397, 2022.
\newblock URL \url{https://api.semanticscholar.org/CorpusID:250920542}.

\bibitem[Chen et~al.(2021)Chen, Tworek, Jun, Yuan, Ponde, Kaplan, Edwards, Burda, Joseph, Brockman, Ray, Puri, Krueger, Petrov, Khlaaf, Sastry, Mishkin, Chan, Gray, Ryder, Pavlov, Power, Kaiser, Bavarian, Winter, Tillet, Such, Cummings, Plappert, Chantzis, Barnes, Herbert-Voss, Guss, Nichol, Babuschkin, Balaji, Jain, Carr, Leike, Achiam, Misra, Morikawa, Radford, Knight, Brundage, Murati, Mayer, Welinder, McGrew, Amodei, McCandlish, Sutskever, and Zaremba]{Chen2021EvaluatingLL}
Chen, M., Tworek, J., Jun, H., Yuan, Q., Ponde, H., Kaplan, J., Edwards, H., Burda, Y., Joseph, N., Brockman, G., Ray, A., Puri, R., Krueger, G., Petrov, M., Khlaaf, H., Sastry, G., Mishkin, P., Chan, B., Gray, S., Ryder, N., Pavlov, M., Power, A., Kaiser, L., Bavarian, M., Winter, C., Tillet, P., Such, F.~P., Cummings, D.~W., Plappert, M., Chantzis, F., Barnes, E., Herbert-Voss, A., Guss, W.~H., Nichol, A., Babuschkin, I., Balaji, S.~A., Jain, S., Carr, A., Leike, J., Achiam, J., Misra, V., Morikawa, E., Radford, A., Knight, M.~M., Brundage, M., Murati, M., Mayer, K., Welinder, P., McGrew, B., Amodei, D., McCandlish, S., Sutskever, I., and Zaremba, W.
\newblock Evaluating large language models trained on code.
\newblock \emph{ArXiv}, abs/2107.03374, 2021.
\newblock URL \url{https://api.semanticscholar.org/CorpusID:235755472}.

\bibitem[Chen et~al.(2023)Chen, Lin, Sch{\"a}rli, and Zhou]{Chen2023TeachingLL}
Chen, X., Lin, M., Sch{\"a}rli, N., and Zhou, D.
\newblock Teaching large language models to self-debug.
\newblock \emph{ArXiv}, abs/2304.05128, 2023.
\newblock URL \url{https://api.semanticscholar.org/CorpusID:258059885}.

\bibitem[Chowdhery et~al.(2022)Chowdhery, Narang, Devlin, Bosma, Mishra, Roberts, Barham, Chung, Sutton, Gehrmann, Schuh, Shi, Tsvyashchenko, Maynez, Rao, Barnes, Tay, Shazeer, Prabhakaran, Reif, Du, Hutchinson, Pope, Bradbury, Austin, Isard, Gur-Ari, Yin, Duke, Levskaya, Ghemawat, Dev, Michalewski, Garc{\'i}a, Misra, Robinson, Fedus, Zhou, Ippolito, Luan, Lim, Zoph, Spiridonov, Sepassi, Dohan, Agrawal, Omernick, Dai, Pillai, Pellat, Lewkowycz, Moreira, Child, Polozov, Lee, Zhou, Wang, Saeta, D{\'i}az, Firat, Catasta, Wei, Meier-Hellstern, Eck, Dean, Petrov, and Fiedel]{Chowdhery2022PaLMSL}
Chowdhery, A., Narang, S., Devlin, J., Bosma, M., Mishra, G., Roberts, A., Barham, P., Chung, H.~W., Sutton, C., Gehrmann, S., Schuh, P., Shi, K., Tsvyashchenko, S., Maynez, J., Rao, A., Barnes, P., Tay, Y., Shazeer, N.~M., Prabhakaran, V., Reif, E., Du, N., Hutchinson, B.~C., Pope, R., Bradbury, J., Austin, J., Isard, M., Gur-Ari, G., Yin, P., Duke, T., Levskaya, A., Ghemawat, S., Dev, S., Michalewski, H., Garc{\'i}a, X., Misra, V., Robinson, K., Fedus, L., Zhou, D., Ippolito, D., Luan, D., Lim, H., Zoph, B., Spiridonov, A., Sepassi, R., Dohan, D., Agrawal, S., Omernick, M., Dai, A.~M., Pillai, T.~S., Pellat, M., Lewkowycz, A., Moreira, E., Child, R., Polozov, O., Lee, K., Zhou, Z., Wang, X., Saeta, B., D{\'i}az, M., Firat, O., Catasta, M., Wei, J., Meier-Hellstern, K.~S., Eck, D., Dean, J., Petrov, S., and Fiedel, N.
\newblock Palm: Scaling language modeling with pathways.
\newblock \emph{ArXiv}, abs/2204.02311, 2022.
\newblock URL \url{https://api.semanticscholar.org/CorpusID:247951931}.

\bibitem[Cobbe et~al.(2021)Cobbe, Kosaraju, Bavarian, Chen, Jun, Kaiser, Plappert, Tworek, Hilton, Nakano, Hesse, and Schulman]{Cobbe2021TrainingVT}
Cobbe, K., Kosaraju, V., Bavarian, M., Chen, M., Jun, H., Kaiser, L., Plappert, M., Tworek, J., Hilton, J., Nakano, R., Hesse, C., and Schulman, J.
\newblock Training verifiers to solve math word problems.
\newblock \emph{ArXiv}, abs/2110.14168, 2021.
\newblock URL \url{https://api.semanticscholar.org/CorpusID:239998651}.

\bibitem[Demeter et~al.(2020)Demeter, Kimmel, and Downey]{demeter-etal-2020-stolen}
Demeter, D., Kimmel, G., and Downey, D.
\newblock Stolen probability: A structural weakness of neural language models.
\newblock In \emph{Proceedings of the 58th Annual Meeting of the Association for Computational Linguistics}, pp.\  2191--2197, Online, July 2020. Association for Computational Linguistics.
\newblock \doi{10.18653/v1/2020.acl-main.198}.
\newblock URL \url{https://aclanthology.org/2020.acl-main.198}.

\bibitem[Devlin et~al.(2019)Devlin, Chang, Lee, and Toutanova]{devlin-etal-2019-bert}
Devlin, J., Chang, M.-W., Lee, K., and Toutanova, K.
\newblock {BERT}: Pre-training of deep bidirectional transformers for language understanding.
\newblock In \emph{Proceedings of the 2019 Conference of the North {A}merican Chapter of the Association for Computational Linguistics: Human Language Technologies, Volume 1 (Long and Short Papers)}, pp.\  4171--4186, Minneapolis, Minnesota, June 2019. Association for Computational Linguistics.
\newblock \doi{10.18653/v1/N19-1423}.
\newblock URL \url{https://aclanthology.org/N19-1423}.

\bibitem[Diao et~al.(2023)Diao, Wang, Lin, and Zhang]{Diao2023ActivePW}
Diao, S., Wang, P., Lin, Y., and Zhang, T.
\newblock Active prompting with chain-of-thought for large language models.
\newblock \emph{ArXiv}, abs/2302.12246, 2023.
\newblock URL \url{https://api.semanticscholar.org/CorpusID:257102707}.

\bibitem[Dong et~al.(2023{\natexlab{a}})Dong, Ding, Jiang, Li, Li, and Jin]{Dong2023CodeScoreEC}
Dong, Y., Ding, J., Jiang, X., Li, Z., Li, G., and Jin, Z.
\newblock Codescore: Evaluating code generation by learning code execution.
\newblock \emph{ArXiv}, abs/2301.09043, 2023{\natexlab{a}}.
\newblock URL \url{https://api.semanticscholar.org/CorpusID:256105296}.

\bibitem[Dong et~al.(2023{\natexlab{b}})Dong, Jiang, Jin, and Li]{Dong2023SelfcollaborationCG}
Dong, Y., Jiang, X., Jin, Z., and Li, G.
\newblock Self-collaboration code generation via chatgpt.
\newblock \emph{ArXiv}, abs/2304.07590, 2023{\natexlab{b}}.
\newblock URL \url{https://api.semanticscholar.org/CorpusID:258179537}.

\bibitem[Du et~al.(2021)Du, Huang, Dai, Tong, Lepikhin, Xu, Krikun, Zhou, Yu, Firat, Zoph, Fedus, Bosma, Zhou, Wang, Wang, Webster, Pellat, Robinson, Meier-Hellstern, Duke, Dixon, Zhang, Le, Wu, Chen, and Cui]{Du2021GLaMES}
Du, N., Huang, Y., Dai, A.~M., Tong, S., Lepikhin, D., Xu, Y., Krikun, M., Zhou, Y., Yu, A.~W., Firat, O., Zoph, B., Fedus, L., Bosma, M., Zhou, Z., Wang, T., Wang, Y.~E., Webster, K., Pellat, M., Robinson, K., Meier-Hellstern, K.~S., Duke, T., Dixon, L., Zhang, K., Le, Q.~V., Wu, Y., Chen, Z., and Cui, C.
\newblock Glam: Efficient scaling of language models with mixture-of-experts.
\newblock \emph{ArXiv}, abs/2112.06905, 2021.
\newblock URL \url{https://api.semanticscholar.org/CorpusID:245124124}.

\bibitem[Fried et~al.(2022)Fried, Aghajanyan, Lin, Wang, Wallace, Shi, Zhong, tau Yih, Zettlemoyer, and Lewis]{Fried2022InCoderAG}
Fried, D., Aghajanyan, A., Lin, J., Wang, S.~I., Wallace, E., Shi, F., Zhong, R., tau Yih, W., Zettlemoyer, L., and Lewis, M.
\newblock Incoder: A generative model for code infilling and synthesis.
\newblock \emph{ArXiv}, abs/2204.05999, 2022.
\newblock URL \url{https://api.semanticscholar.org/CorpusID:248157108}.

\bibitem[Gao et~al.(2023)Gao, Sheng, Xiang, Xiong, Wang, and Zhang]{Gao2023ChatRECTI}
Gao, Y., Sheng, T., Xiang, Y., Xiong, Y., Wang, H., and Zhang, J.
\newblock Chat-rec: Towards interactive and explainable llms-augmented recommender system.
\newblock \emph{ArXiv}, abs/2303.14524, 2023.
\newblock URL \url{https://api.semanticscholar.org/CorpusID:257766541}.

\bibitem[Hao et~al.(2023)Hao, Gu, Ma, Hong, Wang, Wang, and Hu]{Hao2023ReasoningWL}
Hao, S., Gu, Y., Ma, H., Hong, J.~J., Wang, Z., Wang, D.~Z., and Hu, Z.
\newblock Reasoning with language model is planning with world model.
\newblock \emph{ArXiv}, abs/2305.14992, 2023.
\newblock URL \url{https://api.semanticscholar.org/CorpusID:258865812}.

\bibitem[Heafield et~al.(2013)Heafield, Pouzyrevsky, Clark, and Koehn]{heafield-etal-2013-scalable}
Heafield, K., Pouzyrevsky, I., Clark, J.~H., and Koehn, P.
\newblock Scalable modified {K}neser-{N}ey language model estimation.
\newblock In \emph{Proceedings of the 51st Annual Meeting of the Association for Computational Linguistics (Volume 2: Short Papers)}, pp.\  690--696, Sofia, Bulgaria, August 2013. Association for Computational Linguistics.
\newblock URL \url{https://aclanthology.org/P13-2121}.

\bibitem[Hestness et~al.(2017)Hestness, Narang, Ardalani, Diamos, Jun, Kianinejad, Patwary, Yang, and Zhou]{Hestness2017DeepLS}
Hestness, J., Narang, S., Ardalani, N., Diamos, G.~F., Jun, H., Kianinejad, H., Patwary, M. M.~A., Yang, Y., and Zhou, Y.
\newblock Deep learning scaling is predictable, empirically.
\newblock \emph{ArXiv}, abs/1712.00409, 2017.
\newblock URL \url{https://api.semanticscholar.org/CorpusID:2222076}.

\bibitem[Hoffmann et~al.(2022)Hoffmann, Borgeaud, Mensch, Buchatskaya, Cai, Rutherford, de~Las~Casas, Hendricks, Welbl, Clark, Hennigan, Noland, Millican, van~den Driessche, Damoc, Guy, Osindero, Simonyan, Elsen, Rae, Vinyals, and Sifre]{Hoffmann2022TrainingCL}
Hoffmann, J., Borgeaud, S., Mensch, A., Buchatskaya, E., Cai, T., Rutherford, E., de~Las~Casas, D., Hendricks, L.~A., Welbl, J., Clark, A., Hennigan, T., Noland, E., Millican, K., van~den Driessche, G., Damoc, B., Guy, A., Osindero, S., Simonyan, K., Elsen, E., Rae, J.~W., Vinyals, O., and Sifre, L.
\newblock Training compute-optimal large language models.
\newblock \emph{ArXiv}, abs/2203.15556, 2022.
\newblock URL \url{https://api.semanticscholar.org/CorpusID:247778764}.

\bibitem[Huang et~al.(2022)Huang, Gu, Hou, Wu, Wang, Yu, and Han]{Huang2022LargeLM}
Huang, J., Gu, S.~S., Hou, L., Wu, Y., Wang, X., Yu, H., and Han, J.
\newblock Large language models can self-improve.
\newblock \emph{ArXiv}, abs/2210.11610, 2022.
\newblock URL \url{https://api.semanticscholar.org/CorpusID:253080328}.

\bibitem[Inala et~al.(2022)Inala, Wang, Yang, Codas, Encarnaci'on, Lahiri, Musuvathi, and Gao]{Inala2022FaultAwareNC}
Inala, J.~P., Wang, C., Yang, M., Codas, A., Encarnaci'on, M., Lahiri, S.~K., Musuvathi, M., and Gao, J.
\newblock Fault-aware neural code rankers.
\newblock \emph{ArXiv}, abs/2206.03865, 2022.
\newblock URL \url{https://api.semanticscholar.org/CorpusID:249462026}.

\bibitem[Jiang et~al.(2023)Jiang, Dong, Wang, Shang, and Li]{Jiang2023SelfplanningCG}
Jiang, X., Dong, Y., Wang, L., Shang, Q., and Li, G.
\newblock Self-planning code generation with large language model.
\newblock \emph{ArXiv}, abs/2303.06689, 2023.
\newblock URL \url{https://api.semanticscholar.org/CorpusID:257495755}.

\bibitem[Le et~al.(2022)Le, Wang, Gotmare, Savarese, and Hoi]{Le2022CodeRLMC}
Le, H., Wang, Y., Gotmare, A.~D., Savarese, S., and Hoi, S. C.~H.
\newblock Coderl: Mastering code generation through pretrained models and deep reinforcement learning.
\newblock \emph{ArXiv}, abs/2207.01780, 2022.
\newblock URL \url{https://api.semanticscholar.org/CorpusID:250280117}.

\bibitem[Le et~al.(2023)Le, Chen, Saha, Gokul, Sahoo, and Joty]{Le2023CodeChainTM}
Le, H., Chen, H., Saha, A., Gokul, A., Sahoo, D., and Joty, S.~R.
\newblock Codechain: Towards modular code generation through chain of self-revisions with representative sub-modules.
\newblock \emph{ArXiv}, abs/2310.08992, 2023.
\newblock URL \url{https://api.semanticscholar.org/CorpusID:264128082}.

\bibitem[Li et~al.(2023{\natexlab{a}})Li, Li, Li, and Jin]{Li2023StructuredCP}
Li, J., Li, G., Li, Y., and Jin, Z.
\newblock Structured chain-of-thought prompting for code generation.
\newblock 2023{\natexlab{a}}.
\newblock URL \url{https://api.semanticscholar.org/CorpusID:258615421}.

\bibitem[Li et~al.(2023{\natexlab{b}})Li, Zhou, Huang, Chen, and Chen]{Li2023ElicitingTT}
Li, J., Zhou, H., Huang, S., Chen, S., and Chen, J.
\newblock Eliciting the translation ability of large language models via multilingual finetuning with translation instructions.
\newblock \emph{ArXiv}, abs/2305.15083, 2023{\natexlab{b}}.
\newblock URL \url{https://api.semanticscholar.org/CorpusID:258865882}.

\bibitem[Li et~al.(2023{\natexlab{c}})Li, Allal, Zi, Muennighoff, Kocetkov, Mou, Marone, Akiki, Li, Chim, Liu, Zheltonozhskii, Zhuo, Wang, Dehaene, Davaadorj, Lamy-Poirier, Monteiro, Shliazhko, Gontier, Meade, Zebaze, Yee, Umapathi, Zhu, Lipkin, Oblokulov, Wang, Murthy, Stillerman, Patel, Abulkhanov, Zocca, Dey, Zhang, Fahmy, Bhattacharyya, Yu, Singh, Luccioni, Villegas, Kunakov, Zhdanov, Romero, Lee, Timor, Ding, Schlesinger, Schoelkopf, Ebert, Dao, Mishra, Gu, Robinson, Anderson, Dolan-Gavitt, Contractor, Reddy, Fried, Bahdanau, Jernite, Ferrandis, Hughes, Wolf, Guha, von Werra, and de~Vries]{Li2023StarCoderMT}
Li, R., Allal, L.~B., Zi, Y., Muennighoff, N., Kocetkov, D., Mou, C., Marone, M., Akiki, C., Li, J., Chim, J., Liu, Q., Zheltonozhskii, E., Zhuo, T.~Y., Wang, T., Dehaene, O., Davaadorj, M., Lamy-Poirier, J., Monteiro, J., Shliazhko, O., Gontier, N., Meade, N., Zebaze, A., Yee, M.-H., Umapathi, L.~K., Zhu, J., Lipkin, B., Oblokulov, M., Wang, Z., Murthy, R., Stillerman, J., Patel, S.~S., Abulkhanov, D., Zocca, M., Dey, M., Zhang, Z., Fahmy, N., Bhattacharyya, U., Yu, W., Singh, S., Luccioni, S., Villegas, P., Kunakov, M., Zhdanov, F., Romero, M., Lee, T., Timor, N., Ding, J., Schlesinger, C., Schoelkopf, H., Ebert, J., Dao, T., Mishra, M., Gu, A., Robinson, J., Anderson, C.~J., Dolan-Gavitt, B., Contractor, D., Reddy, S., Fried, D., Bahdanau, D., Jernite, Y., Ferrandis, C.~M., Hughes, S.~M., Wolf, T., Guha, A., von Werra, L., and de~Vries, H.
\newblock Starcoder: may the source be with you!
\newblock \emph{ArXiv}, abs/2305.06161, 2023{\natexlab{c}}.
\newblock URL \url{https://api.semanticscholar.org/CorpusID:258588247}.

\bibitem[Li et~al.(2022{\natexlab{a}})Li, Choi, Chung, Kushman, Schrittwieser, Leblond, Tom, Eccles, Keeling, Gimeno, Lago, Hubert, Choy, de, d’Autume, Babuschkin, Chen, Huang, Welbl, Gowal, Alexey, Cherepanov, Molloy, Mankowitz, Robson, Kohli, de, Freitas, Kavukcuoglu, and Vinyals]{Li2022CompetitionlevelCG}
Li, Y., Choi, D.~H., Chung, J., Kushman, N., Schrittwieser, J., Leblond, R., Tom, Eccles, Keeling, J., Gimeno, F., Lago, A.~D., Hubert, T., Choy, P., de, C., d’Autume, M., Babuschkin, I., Chen, X., Huang, P.-S., Welbl, J., Gowal, S., Alexey, Cherepanov, Molloy, J., Mankowitz, D.~J., Robson, E.~S., Kohli, P., de, N., Freitas, Kavukcuoglu, K., and Vinyals, O.
\newblock Competition-level code generation with alphacode.
\newblock \emph{Science}, 378:\penalty0 1092 -- 1097, 2022{\natexlab{a}}.
\newblock URL \url{https://api.semanticscholar.org/CorpusID:246527904}.

\bibitem[Li et~al.(2022{\natexlab{b}})Li, Lin, Zhang, Fu, Chen, Lou, and Chen]{Li2022MakingLM}
Li, Y., Lin, Z., Zhang, S., Fu, Q., Chen, B., Lou, J.-G., and Chen, W.
\newblock Making language models better reasoners with step-aware verifier.
\newblock In \emph{Annual Meeting of the Association for Computational Linguistics}, 2022{\natexlab{b}}.
\newblock URL \url{https://api.semanticscholar.org/CorpusID:259370847}.

\bibitem[Liu et~al.(2022)Liu, Patwary, Prenger, Prabhumoye, Ping, Shoeybi, and Catanzaro]{liu-etal-2022-multi}
Liu, Z., Patwary, M., Prenger, R., Prabhumoye, S., Ping, W., Shoeybi, M., and Catanzaro, B.
\newblock Multi-stage prompting for knowledgeable dialogue generation.
\newblock In \emph{Findings of the Association for Computational Linguistics: ACL 2022}, pp.\  1317--1337, Dublin, Ireland, May 2022. Association for Computational Linguistics.
\newblock \doi{10.18653/v1/2022.findings-acl.104}.
\newblock URL \url{https://aclanthology.org/2022.findings-acl.104}.

\bibitem[Madaan et~al.(2023)Madaan, Tandon, Gupta, Hallinan, Gao, Wiegreffe, Alon, Dziri, Prabhumoye, Yang, Welleck, Majumder, Gupta, Yazdanbakhsh, and Clark]{Madaan2023SelfRefineIR}
Madaan, A., Tandon, N., Gupta, P., Hallinan, S., Gao, L., Wiegreffe, S., Alon, U., Dziri, N., Prabhumoye, S., Yang, Y., Welleck, S., Majumder, B.~P., Gupta, S., Yazdanbakhsh, A., and Clark, P.
\newblock Self-refine: Iterative refinement with self-feedback.
\newblock \emph{ArXiv}, abs/2303.17651, 2023.
\newblock URL \url{https://api.semanticscholar.org/CorpusID:257900871}.

\bibitem[Ni et~al.(2023)Ni, Iyer, Radev, Stoyanov, tau Yih, Wang, and Lin]{Ni2023LEVERLT}
Ni, A., Iyer, S., Radev, D.~R., Stoyanov, V., tau Yih, W., Wang, S.~I., and Lin, X.~V.
\newblock Lever: Learning to verify language-to-code generation with execution.
\newblock \emph{ArXiv}, abs/2302.08468, 2023.
\newblock URL \url{https://api.semanticscholar.org/CorpusID:256900680}.

\bibitem[Nijkamp et~al.(2022)Nijkamp, Pang, Hayashi, Tu, Wang, Zhou, Savarese, and Xiong]{Nijkamp2022CodeGenAO}
Nijkamp, E., Pang, B., Hayashi, H., Tu, L., Wang, H., Zhou, Y., Savarese, S., and Xiong, C.
\newblock Codegen: An open large language model for code with multi-turn program synthesis.
\newblock In \emph{International Conference on Learning Representations}, 2022.
\newblock URL \url{https://api.semanticscholar.org/CorpusID:252668917}.

\bibitem[Olausson et~al.(2023)Olausson, Inala, Wang, Gao, and Solar-Lezama]{Olausson2023IsSA}
Olausson, T.~X., Inala, J.~P., Wang, C., Gao, J., and Solar-Lezama, A.
\newblock Is self-repair a silver bullet for code generation?
\newblock 2023.
\newblock URL \url{https://api.semanticscholar.org/CorpusID:259187989}.

\bibitem[Olmo et~al.(2021)Olmo, Sreedharan, and Kambhampati]{Olmo2021GPT3toplanEP}
Olmo, A., Sreedharan, S., and Kambhampati, S.
\newblock Gpt3-to-plan: Extracting plans from text using gpt-3.
\newblock \emph{ArXiv}, abs/2106.07131, 2021.
\newblock URL \url{https://api.semanticscholar.org/CorpusID:235421645}.

\bibitem[OpenAI(2023)]{OpenAI2023GPT4TR}
OpenAI.
\newblock Gpt-4 technical report.
\newblock \emph{ArXiv}, abs/2303.08774, 2023.
\newblock URL \url{https://api.semanticscholar.org/CorpusID:257532815}.

\bibitem[Pi et~al.(2022)Pi, Liu, Chen, Ziyadi, Lin, Gao, Fu, Lou, and Chen]{Pi2022ReasoningLP}
Pi, X., Liu, Q., Chen, B., Ziyadi, M., Lin, Z., Gao, Y., Fu, Q., Lou, J.-G., and Chen, W.
\newblock Reasoning like program executors.
\newblock \emph{ArXiv}, abs/2201.11473, 2022.
\newblock URL \url{https://api.semanticscholar.org/CorpusID:246294995}.

\bibitem[Radford et~al.(2019)Radford, Wu, Child, Luan, Amodei, and Sutskever]{Radford2019LanguageMA}
Radford, A., Wu, J., Child, R., Luan, D., Amodei, D., and Sutskever, I.
\newblock Language models are unsupervised multitask learners.
\newblock 2019.
\newblock URL \url{https://api.semanticscholar.org/CorpusID:160025533}.

\bibitem[Rae et~al.(2021)Rae, Borgeaud, Cai, Millican, Hoffmann, Song, Aslanides, Henderson, Ring, Young, Rutherford, Hennigan, Menick, Cassirer, Powell, van~den Driessche, Hendricks, Rauh, Huang, Glaese, Welbl, Dathathri, Huang, Uesato, Mellor, Higgins, Creswell, McAleese, Wu, Elsen, Jayakumar, Buchatskaya, Budden, Sutherland, Simonyan, Paganini, Sifre, Martens, Li, Kuncoro, Nematzadeh, Gribovskaya, Donato, Lazaridou, Mensch, Lespiau, Tsimpoukelli, Grigorev, Fritz, Sottiaux, Pajarskas, Pohlen, Gong, Toyama, de~Masson~d'Autume, Li, Terzi, Mikulik, Babuschkin, Clark, de~Las~Casas, Guy, Jones, Bradbury, Johnson, Hechtman, Weidinger, Gabriel, Isaac, Lockhart, Osindero, Rimell, Dyer, Vinyals, Ayoub, Stanway, Bennett, Hassabis, Kavukcuoglu, and Irving]{Rae2021ScalingLM}
Rae, J.~W., Borgeaud, S., Cai, T., Millican, K., Hoffmann, J., Song, F., Aslanides, J., Henderson, S., Ring, R., Young, S., Rutherford, E., Hennigan, T., Menick, J., Cassirer, A., Powell, R., van~den Driessche, G., Hendricks, L.~A., Rauh, M., Huang, P.-S., Glaese, A., Welbl, J., Dathathri, S., Huang, S., Uesato, J., Mellor, J. F.~J., Higgins, I., Creswell, A., McAleese, N., Wu, A., Elsen, E., Jayakumar, S.~M., Buchatskaya, E., Budden, D., Sutherland, E., Simonyan, K., Paganini, M., Sifre, L., Martens, L., Li, X.~L., Kuncoro, A., Nematzadeh, A., Gribovskaya, E., Donato, D., Lazaridou, A., Mensch, A., Lespiau, J.-B., Tsimpoukelli, M., Grigorev, N.~K., Fritz, D., Sottiaux, T., Pajarskas, M., Pohlen, T., Gong, Z., Toyama, D., de~Masson~d'Autume, C., Li, Y., Terzi, T., Mikulik, V., Babuschkin, I., Clark, A., de~Las~Casas, D., Guy, A., Jones, C., Bradbury, J., Johnson, M.~G., Hechtman, B.~A., Weidinger, L., Gabriel, I., Isaac, W.~S., Lockhart, E., Osindero, S., Rimell, L., Dyer, C., Vinyals, O., Ayoub, K.~W.,
  Stanway, J., Bennett, L.~L., Hassabis, D., Kavukcuoglu, K., and Irving, G.
\newblock Scaling language models: Methods, analysis \& insights from training gopher.
\newblock \emph{ArXiv}, abs/2112.11446, 2021.
\newblock URL \url{https://api.semanticscholar.org/CorpusID:245353475}.

\bibitem[Raffel et~al.(2019)Raffel, Shazeer, Roberts, Lee, Narang, Matena, Zhou, Li, and Liu]{Raffel2019ExploringTL}
Raffel, C., Shazeer, N.~M., Roberts, A., Lee, K., Narang, S., Matena, M., Zhou, Y., Li, W., and Liu, P.~J.
\newblock Exploring the limits of transfer learning with a unified text-to-text transformer.
\newblock \emph{ArXiv}, abs/1910.10683, 2019.
\newblock URL \url{https://api.semanticscholar.org/CorpusID:204838007}.

\bibitem[Rosenfeld et~al.(2019)Rosenfeld, Rosenfeld, Belinkov, and Shavit]{Rosenfeld2019ACP}
Rosenfeld, J.~S., Rosenfeld, A., Belinkov, Y., and Shavit, N.
\newblock A constructive prediction of the generalization error across scales.
\newblock \emph{ArXiv}, abs/1909.12673, 2019.
\newblock URL \url{https://api.semanticscholar.org/CorpusID:203592013}.

\bibitem[Shi et~al.(2022)Shi, Fried, Ghazvininejad, Zettlemoyer, and Wang]{Shi2022NaturalLT}
Shi, F., Fried, D., Ghazvininejad, M., Zettlemoyer, L., and Wang, S.~I.
\newblock Natural language to code translation with execution.
\newblock \emph{ArXiv}, abs/2204.11454, 2022.
\newblock URL \url{https://api.semanticscholar.org/CorpusID:248377325}.

\bibitem[Shinn et~al.(2023)Shinn, Cassano, Labash, Gopinath, Narasimhan, and Yao]{Shinn2023ReflexionLA}
Shinn, N., Cassano, F., Labash, B., Gopinath, A., Narasimhan, K., and Yao, S.
\newblock Reflexion: Language agents with verbal reinforcement learning.
\newblock 2023.
\newblock URL \url{https://api.semanticscholar.org/CorpusID:258833055}.

\bibitem[Shoeybi et~al.(2019)Shoeybi, Patwary, Puri, LeGresley, Casper, and Catanzaro]{Shoeybi2019MegatronLMTM}
Shoeybi, M., Patwary, M., Puri, R., LeGresley, P., Casper, J., and Catanzaro, B.
\newblock Megatron-lm: Training multi-billion parameter language models using model parallelism.
\newblock \emph{ArXiv}, abs/1909.08053, 2019.
\newblock URL \url{https://api.semanticscholar.org/CorpusID:202660670}.

\bibitem[Shum et~al.(2023)Shum, Diao, and Zhang]{Shum2023AutomaticPA}
Shum, K., Diao, S., and Zhang, T.
\newblock Automatic prompt augmentation and selection with chain-of-thought from labeled data.
\newblock \emph{ArXiv}, abs/2302.12822, 2023.
\newblock URL \url{https://api.semanticscholar.org/CorpusID:257205763}.

\bibitem[Smith et~al.(2022)Smith, Patwary, Norick, LeGresley, Rajbhandari, Casper, Liu, Prabhumoye, Zerveas, Korthikanti, Zhang, Child, Aminabadi, Bernauer, Song, Shoeybi, He, Houston, Tiwary, and Catanzaro]{Smith2022UsingDA}
Smith, S., Patwary, M., Norick, B., LeGresley, P., Rajbhandari, S., Casper, J., Liu, Z., Prabhumoye, S., Zerveas, G., Korthikanti, V.~A., Zhang, E., Child, R., Aminabadi, R.~Y., Bernauer, J., Song, X., Shoeybi, M., He, Y., Houston, M., Tiwary, S., and Catanzaro, B.
\newblock Using deepspeed and megatron to train megatron-turing nlg 530b, a large-scale generative language model.
\newblock \emph{ArXiv}, abs/2201.11990, 2022.
\newblock URL \url{https://api.semanticscholar.org/CorpusID:246411325}.

\bibitem[Tan et~al.(2021)Tan, Zhang, Wang, and Liu]{Tan2021MSPMP}
Tan, Z., Zhang, X., Wang, S., and Liu, Y.
\newblock Msp: Multi-stage prompting for making pre-trained language models better translators.
\newblock \emph{ArXiv}, abs/2110.06609, 2021.
\newblock URL \url{https://api.semanticscholar.org/CorpusID:238744065}.

\bibitem[Thoppilan et~al.(2022)Thoppilan, Freitas, Hall, Shazeer, Kulshreshtha, Cheng, Jin, Bos, Baker, Du, Li, Lee, Zheng, Ghafouri, Menegali, Huang, Krikun, Lepikhin, Qin, Chen, Xu, Chen, Roberts, Bosma, Zhou, Chang, Krivokon, Rusch, Pickett, Meier-Hellstern, Morris, Doshi, Santos, Duke, S{\o}raker, Zevenbergen, Prabhakaran, D{\'i}az, Hutchinson, Olson, Molina, Hoffman-John, Lee, Aroyo, Rajakumar, Butryna, Lamm, Kuzmina, Fenton, Cohen, Bernstein, Kurzweil, Aguera-Arcas, Cui, Croak, hsin Chi, and Le]{Thoppilan2022LaMDALM}
Thoppilan, R., Freitas, D.~D., Hall, J., Shazeer, N.~M., Kulshreshtha, A., Cheng, H.-T., Jin, A., Bos, T., Baker, L., Du, Y., Li, Y., Lee, H., Zheng, H.~S., Ghafouri, A., Menegali, M., Huang, Y., Krikun, M., Lepikhin, D., Qin, J., Chen, D., Xu, Y., Chen, Z., Roberts, A., Bosma, M., Zhou, Y., Chang, C.-C., Krivokon, I.~A., Rusch, W.~J., Pickett, M., Meier-Hellstern, K.~S., Morris, M.~R., Doshi, T., Santos, R.~D., Duke, T., S{\o}raker, J.~H., Zevenbergen, B., Prabhakaran, V., D{\'i}az, M., Hutchinson, B., Olson, K., Molina, A., Hoffman-John, E., Lee, J., Aroyo, L., Rajakumar, R., Butryna, A., Lamm, M., Kuzmina, V.~O., Fenton, J., Cohen, A., Bernstein, R., Kurzweil, R., Aguera-Arcas, B., Cui, C., Croak, M.~R., hsin Chi, E.~H., and Le, Q.
\newblock Lamda: Language models for dialog applications.
\newblock \emph{ArXiv}, abs/2201.08239, 2022.
\newblock URL \url{https://api.semanticscholar.org/CorpusID:246063428}.

\bibitem[Wang et~al.(2023)Wang, Liu, Wang, Cui, Ding, Liu, and Yu]{Wang2023INTERVENORPT}
Wang, H., Liu, Z., Wang, S., Cui, G., Ding, N., Liu, Z., and Yu, G.
\newblock Intervenor: Prompt the coding ability of large language models with the interactive chain of repairing.
\newblock \emph{ArXiv}, abs/2311.09868, 2023.
\newblock URL \url{https://api.semanticscholar.org/CorpusID:265221349}.

\bibitem[Wang et~al.(2022)Wang, Wei, Schuurmans, Le, hsin Chi, and Zhou]{Wang2022SelfConsistencyIC}
Wang, X., Wei, J., Schuurmans, D., Le, Q., hsin Chi, E.~H., and Zhou, D.
\newblock Self-consistency improves chain of thought reasoning in language models.
\newblock \emph{ArXiv}, abs/2203.11171, 2022.
\newblock URL \url{https://api.semanticscholar.org/CorpusID:247595263}.

\bibitem[Wei et~al.(2022)Wei, Wang, Schuurmans, Bosma, hsin Chi, Xia, Le, and Zhou]{Wei2022ChainOT}
Wei, J., Wang, X., Schuurmans, D., Bosma, M., hsin Chi, E.~H., Xia, F., Le, Q., and Zhou, D.
\newblock Chain of thought prompting elicits reasoning in large language models.
\newblock \emph{ArXiv}, abs/2201.11903, 2022.
\newblock URL \url{https://api.semanticscholar.org/CorpusID:246411621}.

\bibitem[Welleck et~al.(2022)Welleck, Lu, West, Brahman, Shen, Khashabi, and Choi]{Welleck2022GeneratingSB}
Welleck, S., Lu, X., West, P., Brahman, F., Shen, T., Khashabi, D., and Choi, Y.
\newblock Generating sequences by learning to self-correct.
\newblock \emph{ArXiv}, abs/2211.00053, 2022.
\newblock URL \url{https://api.semanticscholar.org/CorpusID:253244506}.

\bibitem[Yao et~al.(2022)Yao, Zhao, Yu, Du, Shafran, Narasimhan, and Cao]{Yao2022ReActSR}
Yao, S., Zhao, J., Yu, D., Du, N., Shafran, I., Narasimhan, K., and Cao, Y.
\newblock React: Synergizing reasoning and acting in language models.
\newblock \emph{ArXiv}, abs/2210.03629, 2022.
\newblock URL \url{https://api.semanticscholar.org/CorpusID:252762395}.

\bibitem[Yao et~al.(2023)Yao, Yu, Zhao, Shafran, Griffiths, Cao, and Narasimhan]{Yao2023TreeOT}
Yao, S., Yu, D., Zhao, J., Shafran, I., Griffiths, T.~L., Cao, Y., and Narasimhan, K.
\newblock Tree of thoughts: Deliberate problem solving with large language models.
\newblock \emph{ArXiv}, abs/2305.10601, 2023.
\newblock URL \url{https://api.semanticscholar.org/CorpusID:258762525}.

\bibitem[Zelikman et~al.(2022)Zelikman, Wu, and Goodman]{Zelikman2022STaRBR}
Zelikman, E., Wu, Y., and Goodman, N.~D.
\newblock Star: Bootstrapping reasoning with reasoning.
\newblock \emph{ArXiv}, abs/2203.14465, 2022.
\newblock URL \url{https://api.semanticscholar.org/CorpusID:247762790}.

\bibitem[Zhang et~al.(2023)Zhang, Li, Li, Li, and Jin]{Zhang2023SelfEditFC}
Zhang, K., Li, Z., Li, J., Li, G., and Jin, Z.
\newblock Self-edit: Fault-aware code editor for code generation.
\newblock \emph{ArXiv}, abs/2305.04087, 2023.
\newblock URL \url{https://api.semanticscholar.org/CorpusID:258557186}.

\bibitem[Zhang et~al.(2022{\natexlab{a}})Zhang, Roller, Goyal, Artetxe, Chen, Chen, Dewan, Diab, Li, Lin, Mihaylov, Ott, Shleifer, Shuster, Simig, Koura, Sridhar, Wang, and Zettlemoyer]{Zhang2022OPTOP}
Zhang, S., Roller, S., Goyal, N., Artetxe, M., Chen, M., Chen, S., Dewan, C., Diab, M.~T., Li, X., Lin, X.~V., Mihaylov, T., Ott, M., Shleifer, S., Shuster, K., Simig, D., Koura, P.~S., Sridhar, A., Wang, T., and Zettlemoyer, L.
\newblock Opt: Open pre-trained transformer language models.
\newblock \emph{ArXiv}, abs/2205.01068, 2022{\natexlab{a}}.
\newblock URL \url{https://api.semanticscholar.org/CorpusID:248496292}.

\bibitem[Zhang et~al.(2022{\natexlab{b}})Zhang, Yu, Hashimoto, Lewis, tau Yih, Fried, and Wang]{Zhang2022CoderRR}
Zhang, T., Yu, T., Hashimoto, T., Lewis, M., tau Yih, W., Fried, D., and Wang, S.~I.
\newblock Coder reviewer reranking for code generation.
\newblock \emph{ArXiv}, abs/2211.16490, 2022{\natexlab{b}}.
\newblock URL \url{https://api.semanticscholar.org/CorpusID:254069951}.

\bibitem[Zhang et~al.(2022{\natexlab{c}})Zhang, Zhang, Li, and Smola]{Zhang2022AutomaticCO}
Zhang, Z., Zhang, A., Li, M., and Smola, A.~J.
\newblock Automatic chain of thought prompting in large language models.
\newblock \emph{ArXiv}, abs/2210.03493, 2022{\natexlab{c}}.
\newblock URL \url{https://api.semanticscholar.org/CorpusID:252762275}.

\bibitem[Zheng et~al.(2023)Zheng, Xia, Zou, Dong, Wang, Xue, Wang, Shen, Wang, Li, Su, Yang, and Tang]{Zheng2023CodeGeeXAP}
Zheng, Q., Xia, X., Zou, X., Dong, Y., Wang, S., Xue, Y., Wang, Z.-Y., Shen, L., Wang, A., Li, Y., Su, T., Yang, Z., and Tang, J.
\newblock Codegeex: A pre-trained model for code generation with multilingual evaluations on humaneval-x.
\newblock \emph{ArXiv}, abs/2303.17568, 2023.
\newblock URL \url{https://api.semanticscholar.org/CorpusID:257834177}.

\bibitem[Zhou et~al.(2022)Zhou, Scharli, Hou, Wei, Scales, Wang, Schuurmans, Bousquet, Le, and hsin Chi]{Zhou2022LeasttoMostPE}
Zhou, D., Scharli, N., Hou, L., Wei, J., Scales, N., Wang, X., Schuurmans, D., Bousquet, O., Le, Q., and hsin Chi, E.~H.
\newblock Least-to-most prompting enables complex reasoning in large language models.
\newblock \emph{ArXiv}, abs/2205.10625, 2022.
\newblock URL \url{https://api.semanticscholar.org/CorpusID:248986239}.

\bibitem[Zhou et~al.(2023)Zhou, Jiang, Cotterell, and Sachan]{Zhou2023EfficientPV}
Zhou, W., Jiang, Y., Cotterell, R., and Sachan, M.
\newblock Efficient prompting via dynamic in-context learning.
\newblock \emph{ArXiv}, abs/2305.11170, 2023.
\newblock URL \url{https://api.semanticscholar.org/CorpusID:258762345}.

\end{thebibliography}
\bibliographystyle{icml2024}

\newpage
\appendix
\onecolumn
\newpage
\appendix

\section{Appendix}\label{sec:appendix}

\subsection{Effectiveness of \xxx with GPT-4}
To illustrate the effectiveness of \xxx with GPT-4, we illustrate the evaluation results in~\cref{tab:gpt4}, we can find that compared to GPT-4, \xxx can improve the effectiveness of code generation in four datasets. For example, CodeCoT increases pass@1 from 67.0\%\footnote{Reported by GPT-4 technical report} to 87.20\% for the HumanEval dataset.

\begin{table*}[h]
    \centering
    \begin{tabular}{c|ccccc}
    \toprule
    Models&HumanEval&HumanEval-ET&MBPP&MBPP-ET\\
    \midrule
    GPT-4&67.0&50.7&68.1&49.2\\
    \xxx + GPT-4&87.20&76.83&92.61&68.87\\
         \bottomrule
    \end{tabular}
    \caption{Evaluation results of \xxx with GPT4. The result of GPT4 is reported on the GPT-4 technical report.}
    \label{tab:gpt4}
\end{table*}

\subsection{Refinement with test case}
As shown in~\cref{fig:pipeline}, we focus on the refinement of code snippets generated by the code generation model, while test cases will not be considered to refine in this process. However, as shown in~\cref{tab:testcase}, we can find that the test case effectiveness can be improved since there is a gap between the 100\% accuracy~(pass@1 in \textbf{canonical\_solution}) and current results. In this section, we try to require the model also refine the test cases during the self-examination procedure.

The evaluation results are shown in~\cref{tab:testcase_refinement} and~\cref{tab:refine_code_test_case}. We can find that first, the test case accuracy in the \textbf{canonical\_solution} will largely increases when we also require the code generation model refine its test cases. For example, the accuracy increases from 67.1\% and 79.0\% to 84.1\% and 93.4\% compared in HumanEval and MBPP datasets. Then, we can also notice that the pass@1 also increases from 79.3\% and 89.5\% to 82.9\% and 90.7\%. Notably, once we add the test cases into self-examination refinement process, we can find the ET datasets' pass@1 largely increased. For example, for the 5 steps, we can find that the pass@1 increases from 69.5\% and 63.0\% to 74.4 and 70.0\% in HumanEval-ET and MBPP-ET datasets.

\begin{table}[h]
    \centering
    \begin{tabular}{c|cc}
    \toprule
         Refine Step&HumanEval&MBPP  \\
         \midrule
         0&67.1&79.0\\
         1&73.2&80.9 \\
         3&79.9&85.2 \\
         5&84.1&93.4 \\
         \bottomrule
    \end{tabular}
    \caption{CodeCoT generated test case's effectiveness in \textbf{canonical\_solution}.}
    \label{tab:testcase_refinement}
\end{table}

\begin{table*}[h]
    \centering
    \begin{tabular}{c|cccc}
    \toprule
         Refine Step&HumanEval&HumanEval-ET&MBPP&MBPP-ET  \\
         \midrule
         1 step& 71.3&60.4&81.7&58.4\\
         1 step*&73.2&64.0&82.5&61.5\\
         3 step& 76.8&67.1&88.3&62.7\\
         3 step*&80.5&72.0&89.5&68.1\\
         5 step&79.3&69.5&89.5&63.0\\
         5 step*&82.9&74.4&90.7&70.0\\
    \bottomrule
    \end{tabular}
    \caption{evaluation results of pass@1 with different self-examination steps for both code snippets and test cases. We use x step means evaluation results that only refine code snippets with x steps. While x step* means both code snippets and test cases will be refined x steps.}
    \label{tab:refine_code_test_case}
\end{table*}

\subsection{Test Cases' Coverage Discussion}
In this section, we discuss whether \xxx can cover more corner cases compared with our baseline strategies which also require LLMs to generate test cases to guide code generation. The evaluation results are shown in~\cref{tab:coverage}, where we compare \xxx with the original ChatGPT generation and CodeT results. For ease of discussion, we require each strategy to generate five test cases to analyze the code line coverage in the \textbf{canonical\_solution}. We can find that \xxx obtains SOTA performance in HumanEval and MBPP datasets. Specifically, \xxx obtains 74.7\% and 79.3\% code line coverage while CodeT only obtains 67.1\% and 73.5\% code line coverage, which indicates that \xxx can cover more corner cases in the \textbf{canonical\_solution}.

\begin{table}[h]
    \centering
    \begin{tabular}{c|cc}
    \toprule
    Models&HumanEval&MBPP\\
    ChatGPT&67.1&58.4\\
    CodeT&67.1&73.5\\
    \xxx&74.7&79.3\\
         \bottomrule
    \end{tabular}
    \caption{Evaluation results for coverage of the generated test cases. In our experiment, we calculate the code line coverage with the first five test cases provided by the tester for each function.}
    \label{tab:coverage}
\end{table}

\subsection{Case Illustration for CodeCoT}

\begin{figure*}
    \centering
    \includegraphics{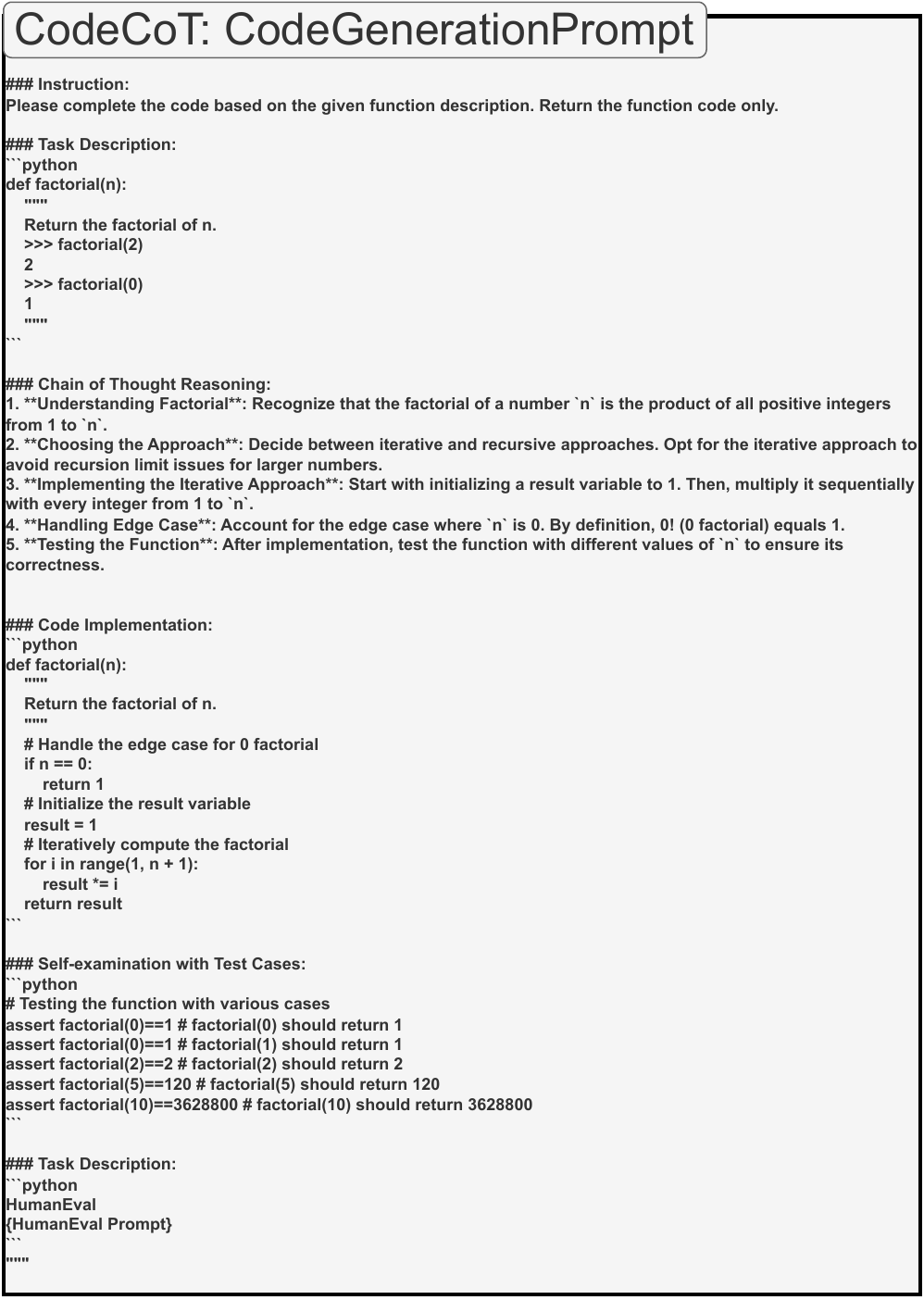}
    \caption{Illustration of CodeCoT code generation prompt template.}
    \label{fig:codegenerationprompt}
\end{figure*}

\begin{figure*}
    \centering
    \includegraphics[width=\textwidth]{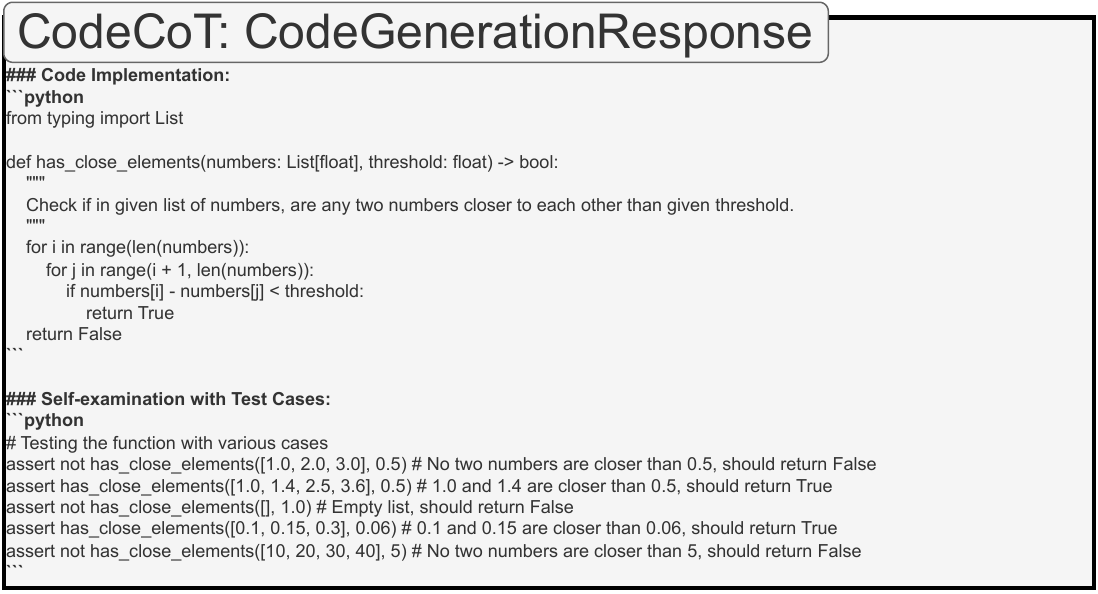}
    \caption{Illustration of CodeCoT code generation response template.}
    \label{fig:enter-label}
\end{figure*}

\begin{figure*}
    \centering
    \includegraphics[width=0.8\textwidth]{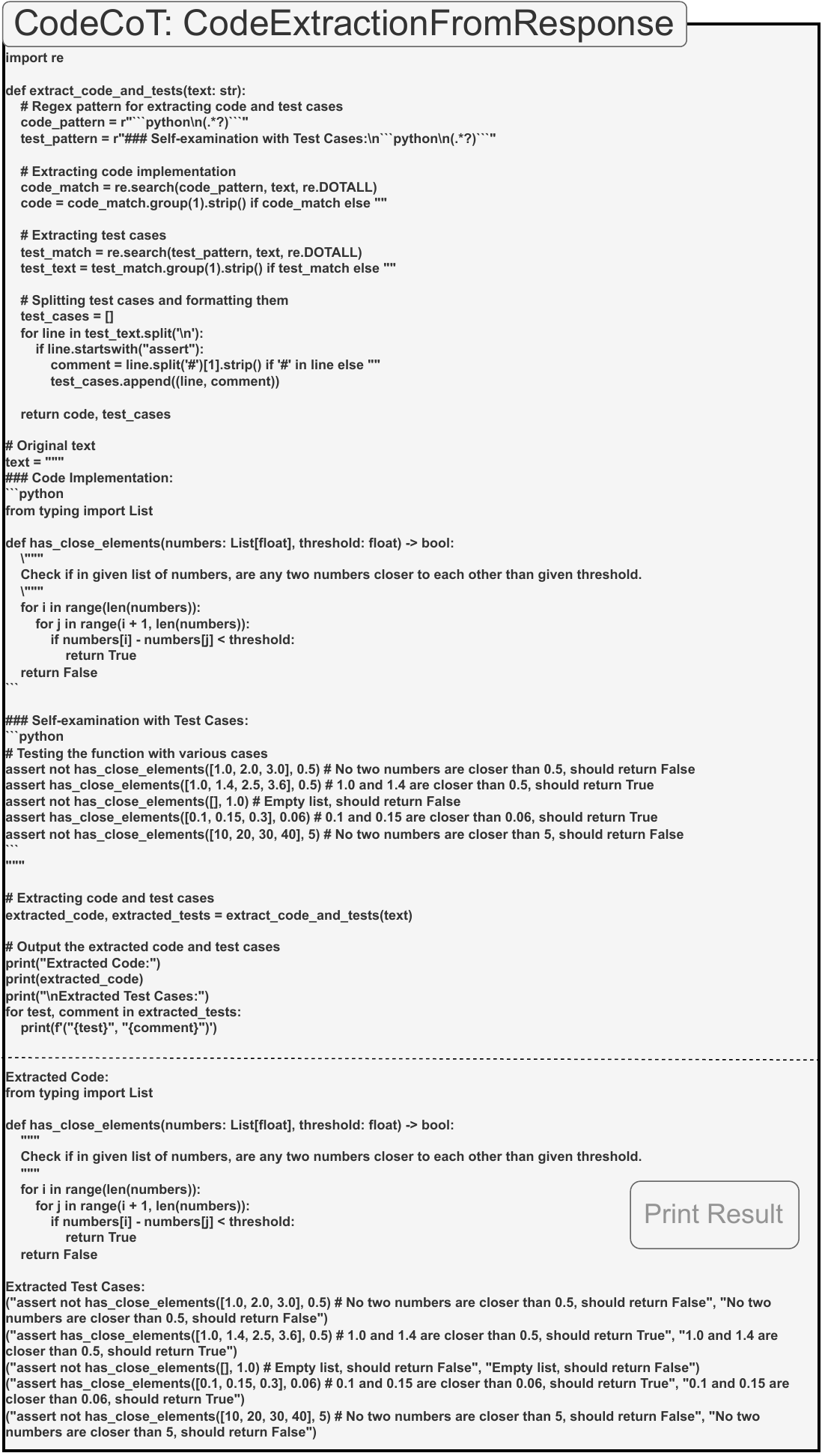}
    \caption{Illustration of CodeCoT response preprocess template.}
    \label{fig:enter-label}
\end{figure*}

\begin{figure*}
    \centering
    \includegraphics[width=\textwidth]{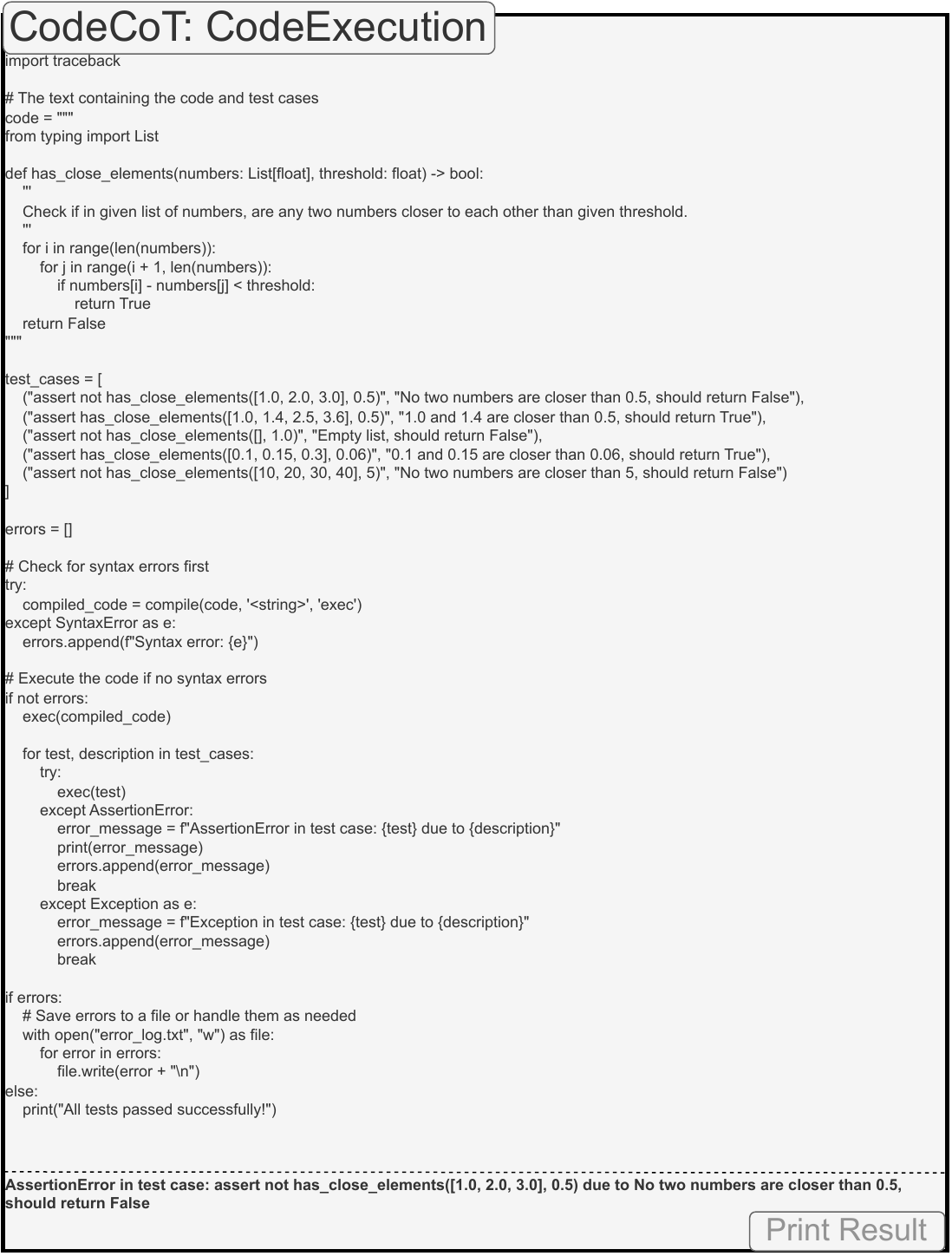}
    \caption{Illustration of CodeCoT Execution template.}
    \label{fig:enter-label}
\end{figure*}

\begin{figure*}
    \centering
    \includegraphics[width=\textwidth]{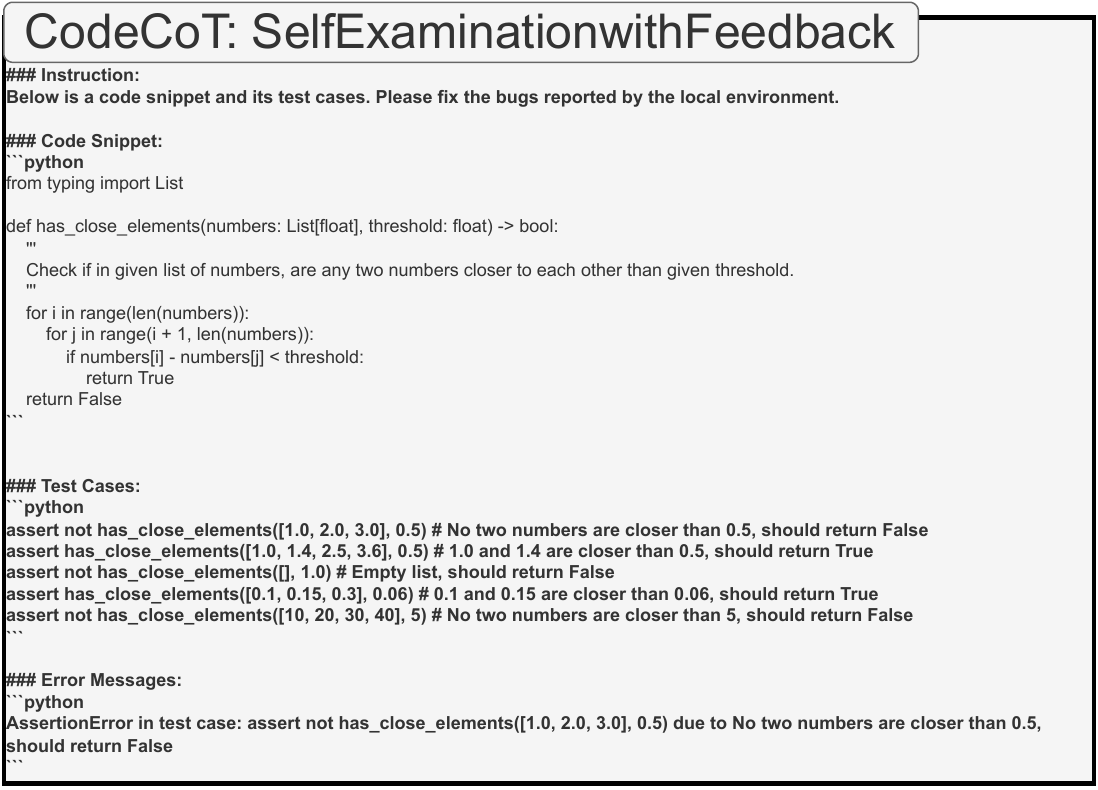}
    \caption{Illustration of CodeCoT Self-examination prompt template.}
    \label{fig:enter-label}
\end{figure*}

\begin{figure*}
    \centering
    \includegraphics[width=\textwidth]{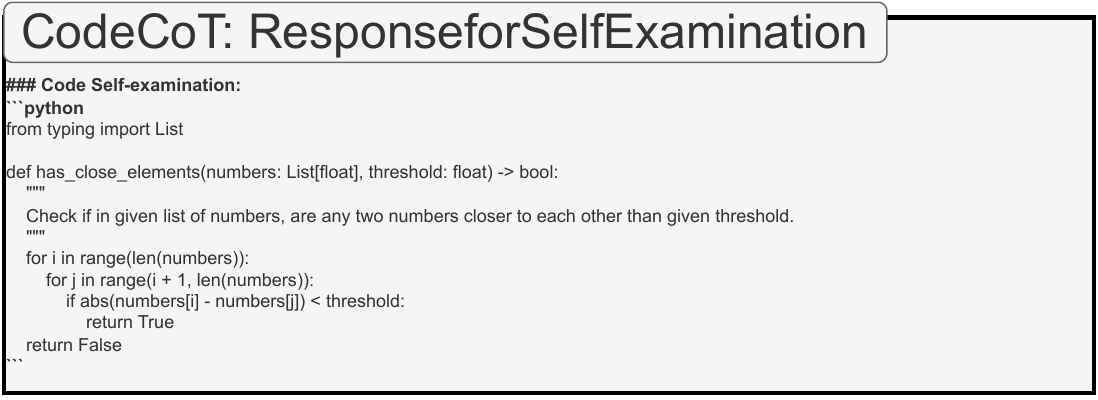}
    \caption{Illustration of CodeCoT Self-examination response template.}
    \label{fig:enter-label}
\end{figure*}

\end{document}